\documentclass{article}

\usepackage[final]{neurips_2021}
\usepackage{graphicx}
\usepackage{tabularx} 
\usepackage{adjustbox}
\usepackage{dcolumn}
\usepackage{amsmath}

\usepackage{array}

\usepackage[T1]{fontenc}    % use 8-bit T1 fonts
\usepackage{hyperref}       % hyperlinks
\usepackage{url}            % simple URL typesetting
\usepackage{booktabs}       % professional-quality tables
\usepackage{amsfonts}       % blackboard math symbols
\usepackage{nicefrac}       % compact symbols for 1/2, etc.
\usepackage{microtype}      % microtypography
\usepackage{xcolor}         % colors
\usepackage{subfig}
\usepackage{multirow}
\usepackage{array}

\title{An Improved Hybrid Recommender System: Integrating Document Context-Based and Behavior-Based Methods}

\author{	
	Meysam Varasteh\\
	University of Tehran\\
	Tehran, Iran \\
	\texttt{meysamvaraste@ut.ac.ir} \\
	\And
	Mehdi Soleiman Nejad \\
	University of Tehran\\
	Tehran, Iran  \\
	\texttt{m.soleimannejad@ut.ac.ir} 	
	\And
	Hadi Moradi \\
	University of Tehran\\
	Tehran, Iran  \\
	\texttt{moradih@ut.ac.ir} 
	\AND
	Mohammad Amin Sadeghi\\
	University of Tehran\\
	Tehran, Iran  \\
	\texttt{asadeghi@ut.ac.ir} 	\\
	\And
	Ahmad Kalhor\\
	University of Tehran\\
	Tehran, Iran  \\
	\texttt{akalhor@ut.ac.ir} 
}

\begin{document}
	\maketitle
	\begin{abstract}
	One of the main challenges in recommender systems is data sparsity which leads to high variance. Several attempts have been made to improve the bias-variance trade-off using auxiliary information. In particular, document modeling-based methods have improved the model's accuracy by using textual data such as reviews, abstracts, and storylines when the user-to-item rating matrix is sparse. However, such models are insufficient to learn optimal representation for users and items. User-based and item-based collaborative filtering, owing to their efficiency and interpretability, have been long used for building recommender systems. They create a profile for each user and item respectively as their historically interacted items and the users who interacted with the target item. 
	\newline
	This work combines these two approaches with document context-aware recommender systems by considering users' opinions on these items. Another advantage of our model is that it supports online personalization. If a user has new interactions, it needs to refresh the user and item history representation vectors instead of updating model parameters. The proposed algorithm is implemented and tested on three real-world datasets that demonstrate our model's effectiveness over the baseline methods. 
	\end{abstract}
	
	\section{Introduction}\label{section:introduction}
	Nowadays, recommender systems have become an integral part of our lives; with the ever-increasing growth of information, these systems guide so many aspects of our life. Recommender systems provide important and relevant cases and filter non-relevant ones, which help us in making good decisions and saving our time.
	One of the main objectives of recommender systems is to model user preferences for items based on the recorded information\cite{koren2009matrix}. User preferences can be extracted through their ratings, clicks, or percentage of views\citep{nakhli2019movie,hu2008collaborative,xin2019cfm}. In most online services, customers can submit their reviews for products and share their opinions with other customers to help them. Researches show that nearly one-third of online shoppers refuse to purchase products that have not received positive feedback from customers; Therefore, it is a mutual benefit for users and the company, which simultaneously increases user's satisfaction and corporate profit\citep{adomavicius2011context,utz2012consumers,von2018influence}. Researchers have recently been using this valuable information to represent users and items better and handle the sparsity problem\citep{10.1145/2645710.2645728,mcauley2013hidden}.
	 \newline
	Various document-based modeling approaches such as Latent Dirichlet Allocation (LDA) and Stacked Denoising Auto-Encoder (SDAE) have been proposed to improve the accuracy of recommender systems by utilizing textual data such as reviews and storylines\citep{wang2015collaborative,wang2011collaborative,mcauley2013hidden,10.1145/2645710.2645728}.
	In addition, some methods have been proposed by integrating the aforementioned topic modeling methods with Collaborative Filtering, known as Collaborative Topic Regression (CTR)\citep{wang2015collaborative,mnih2008probabilistic}. However, such integrated approaches do not fully capture document information. To address this issue, Some works like \citep{zhang2018integrating,zhang2020extreme,kim2016convolutional} utilize Convolutional Neural
	Networks (CNN). CNNs facilitate a deeper understanding of documents and generate a better latent vector than topic modeling methods.
	\newline
	Despite the prevalence and effectiveness of Document Context-Aware models in recommendation systems, we argue that such models are insufficient to learn optimal representation for users and items. The major limitation is that their historical behaviors did not influence users’ interests because their current interests are intrinsically dynamic. The second challenge is how to combine the interaction information of other users concerning their opinions with textual data. To provide better representation that contains both interaction and textual information.
	\newline
	To address these two limitations, we aim to build a Hybrid Recommender System. We create a profile for each user and characterize the user with their interaction history information. The opinions of users on items can capture users' preferences on items that provide better representation. 
	Likewise, We build a profile for each item based on the set of users who have interacted with the target item; even users might express different opinions for each item. By combining these pieces of information with the textual description, we can capture the characteristics of the item from different perspectives.
	\newline
	Another advantage of our model is that it supports online personalization. For online personalization, if a user makes new interactions, the recommender model needs to refresh the top-K recommendation for the user instantaneously, which needs to retrain the model parameters\citep{he2018nais,he2016fast}. However, it takes too many computation resources to perform model retraining in real-time. Instead of updating the parameters, our model can refresh the user and item history representation vector without updating any model parameters.
	\newline
	The reminder of the paper is organized as follows. In section \ref{section:related work} we briefly review the related works. In Section \ref{section:model} we describe our proposed model. In Section \ref{section:experiment} we evaluate our model on three real-world datasets. Finally, in section \ref{section:conclusion} we conclude our work with future directions.

	\section{Related Work}\label{section:related work}
	In this section, we briefly review several closely related works, including general recommendation methods, context-aware recommender systems, and attention mechanism.
	\subsection{General Recommendation}\label{section:General Recommendation}
	Collaborative Filtering (CF) is one of the most popular and widely used filtering methods; the motivation behind it comes from the idea that our future behavior depends on past information to some extent \cite{koren2009matrix}. Also, we often get the best recommendations from other users that have similar tastes to us \cite{cheng2016wide}. 
	CF can be divided into model-based and memory-based approaches\cite{sarwar2001item,deshpande2004item}. Matrix Factorization (MF) is a class of the CF algorithm. The idea behind MF is to represent each user and item in lower dimension latent space, and the objective is to exploit the relationship between users and items latent vectors\citep{koren2009matrix,wei2017collaborative}. By multiplying these vectors, a user's preference, $U$, to an item, $I$, is obtained. For example, if we have $N$ users, $M$ items  and rating matrix $R\subset \mathbb{R}^{N \times M}$, The user and item latent vector respectively  are $u_{i} \subset \mathbb{R}^{k}$ and $v_{j} \subset \mathbb{R}^{k}$ where  $k$ is the latent space size. The predicted rating of $U$ to $I$ is calculated by multiplying the corresponding vectors. 
	\newline
	The other two most common forms in CF are user-based and item-based CF. The user-based collaborative filtering (UCF) is based on looking for users with similar tastes to the active user and representing each item with users who have chosen the target item. In contrast, item-based collaborative filtering (ICF) represents each user with their interacted items and recommends items that are similar to the user profile\citep{xue2019deep,he2018nais}. This paper aims to integrate them with a context-aware recommender system (CARS) that uses the textual description of items as additional information, where items latent vectors contain both contextual and historical interaction information.
	\newline
	In recent years, deep learning, due to its capability of approximating any continuous function and capturing intricate patterns, has garnered attention in many fields, including recommender systems\citep{zhang2019deep,mu2018survey}. Some deep learning methods are used to estimate user preferences. For example, \citep{he2017neural} fused CF with neural architecture, which considers both the linearity of MF and the non-linearity of neural architecture to enhance ranking performance. Another line of work uses auxiliary information such as text, image, and acoustic features in the CF model, which we elaborate in the following subsection.
	
	\subsection{Context-Aware Recommender Systems}\label{section:CARSs}
	The goal of context-aware recommender systems (CARSs) is to model users' preferences by using contextual information in the recommendation process. In CARSs, contextual factors are considered when modeling user profiles or item profiles. Some researches show that adding contextual information to the recommendation process can improve the accuracy and address the sparsity problem.  \citep{kim2016convolutional,zhang2020extreme}. 
	\newline
	  One of the most important and widely used tools in text processing is CNN which has become the cornerstone of deep learning \citep{zhang2018integrating,zhang2020extreme,kim2016convolutional}. Taking advantage of this success, \citep{kim2016convolutional} uses CNN architecture to capture contextual information of the movie's description and combines it with probabilistic matrix factorization (PMF) to enhance rating prediction\citep{mnih2007probabilistic}. In recent years, another one of the stunning successes in deep learning was the attention mechanism, which will be discussed in more detail in section \ref{section:attention}. \citep{zhang2018integrating,zhang2020extreme} have used this mechanism and integrated it with residual networks.\citep{smirnova2017contextual, livne2019deep} proposed a context-aware session-based RecSys by utilizing conditional RNNs, which injects contextual information into input and output layers. It modifies the behavior of the RNN by combining context embedding with item embedding.
	  \newline
	 To address the sparsity problem and improve the accuracy of the CARS, \cite{livne2019deep} proposed a model which uses the encoding-decoding process to learn latent context representations and utilizes sequences of user data derived from contextual conditions.

	\subsection{Attention Mechanism}\label{section:attention}
	The attention in deep learning is based on the screening ability of human beings, which has been applied in many tasks such as Information Retrieval (IR), Natural Language Processing (NLP) and recommender systems\cite{chaudhari2019attentive}. Recommender systems employ the attention mechanism to improve performance and interpretability\cite{sun2019bert4rec}. The basic idea behind this mechanism is to assign different weights to different parts. It can be regarded as a weights vector; 
	Calculate a weight for each feature vector that shows the importance of the corresponding feature, then multiply each of the estimated weights into corresponding inputs\cite{xiao2017attentional}.
	\newline
	The attention model calculates attention weights by a feedforward neural network which is denoted by $\alpha(t,1) $, $\alpha(t,2)$,...,$\alpha(t,t)$. The output of the attention layer is generated using the weighted sum of annotations:
	\begin{eqnarray}
		c_{i}&=&\sum_{j=1}^{T_{x}}\alpha _{ij}h_{j} \\
		\alpha_{ij}&=&\frac{\exp(f(e_{ij}))}{\sum_{k=1}^{T_{x}}\exp(f(e_{ik}))} \\
		f(e_{ij}) &=& h^{T} Relu(W \times e_{i,j}+b)
	\end{eqnarray}
	\section{Our Model} \label{section:model}	

	\subsection{Problem Statement}
	Let $U=\{u_{1},u_{2},...,u_{n}\}$ and $V=\{v_{1},v_{2},...,v_{n}\}$ be the sets of users and items respectively where n is the number of users and m is the number of items. In recommender systems, the user-item rating matrix is denoted by R $\subset \mathbb{R}^{n \times m}$ where each row represents a user $u_{j}$ with $1 \leq j \leq m $ and each column represents an item $i_{k}$ with $1 \leq k \leq n $ . The elements of this matrix are the ratings that are given to items by users. An example input of our model is illustrated as follows:
	\newline
	\begin{eqnarray*}
		\underbrace{[u_{1}^k,u_{2}^k,...,u_{n^{'}}^k]}_\textrm{history of items for $u^{k}$} ~~~~~~~ \underbrace{[w_{1}, w_{2},..., w_{k^{'}}]}_\textrm{words vector}~~~~~~~ \underbrace{[v_{1}^k, v_{2}^k,..., v_{m^{'}}^k]}_\textrm{history of users for $v^k$}
	\end{eqnarray*}
Where $n^{'}<n$, $k^{'}<k$ and $m^{'}<m$. Given the above inputs, our model aims to predict the rating.
	\subsection{General Framework}
\begin{figure}
		\centering
		\includegraphics[width=14cm,height=10cm]{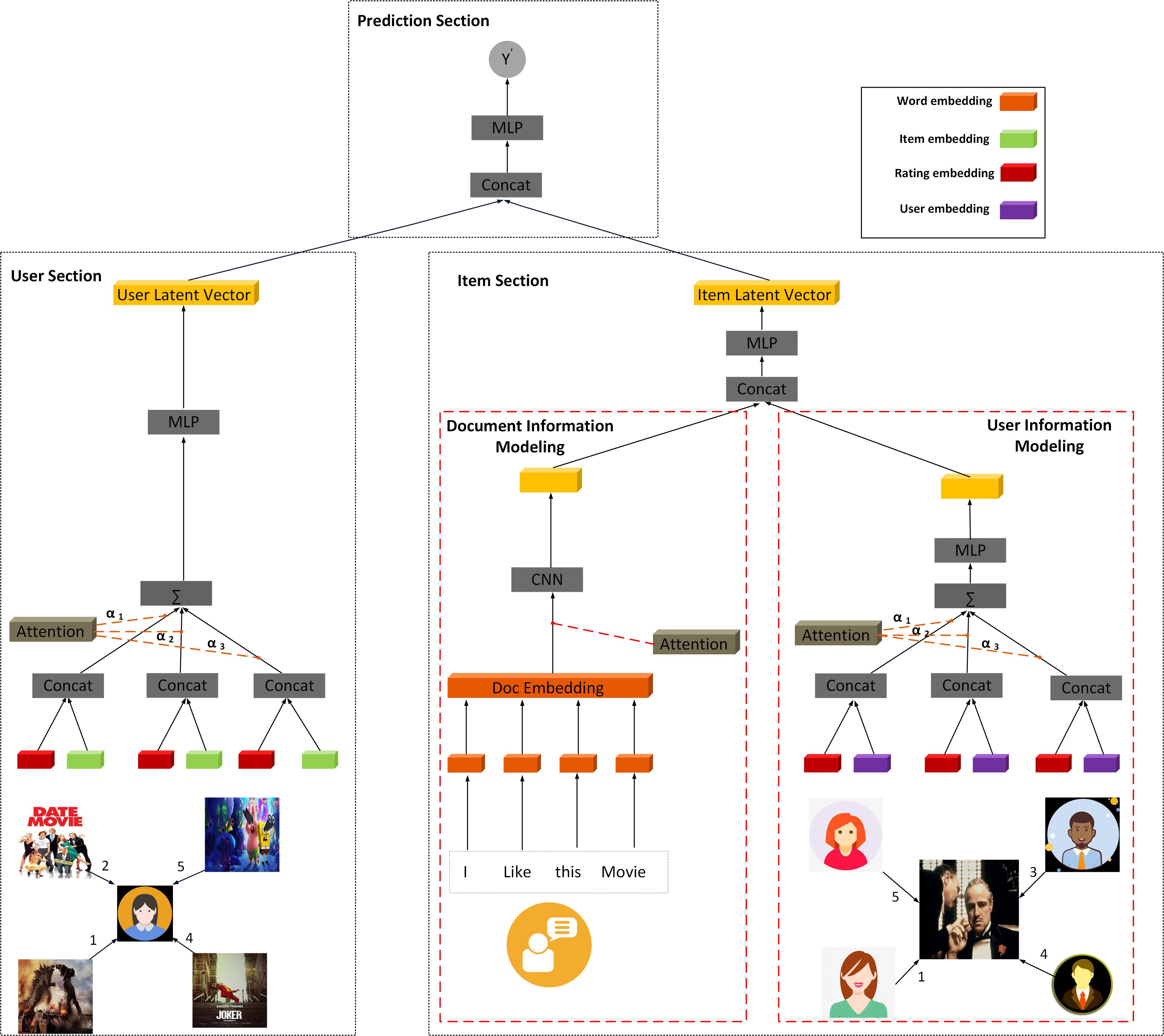}
		\caption{The overall architecture of our proposed model.}
		\label{figure=overal struct}
	\end{figure}

	Fig \ref{figure=overal struct} shows the architecture of our proposed model, which consists of three sections: user section, item section, and rating prediction section. The main objective in the user and item section is to learn a latent vector for each item and user by using available information. The third part is used to predict the rating value of users to items.
	\subsection{User Section}
	The user section aims to learn the user latent vector by using past interactions and opinions of the user. Each user is mapped to a multi-hot vector composed of movies that the user has rated. In this section, two cases are studied: 1) Without rating information 2) With rating information.
	\newline
	\textbf{Without Rating Information}: The input vector as a multi-hot vector contains just the item's id without considering rating information. Then each of the input vectors is passed through the attention network to calculate attention weights, and each of the estimated weights is multiplied by the corresponding vectors.
	\newline
	\textbf{With Rating Information}:  A user can express their satisfaction with items, denoted as a rating score. We can use this auxiliary information as \cite{fan2019graph} and combine it with the user's interaction information to provide a richer user latent vector than baseline methods. Like the previous case, each input vector is passed through the attention network to calculate weights. By multiplying each of the input vectors to corresponding weights, the latent user vector is obtained.
	\subsection{Item Section}
\begin{figure}
		\centering
		\includegraphics[width=14cm,height=8cm]{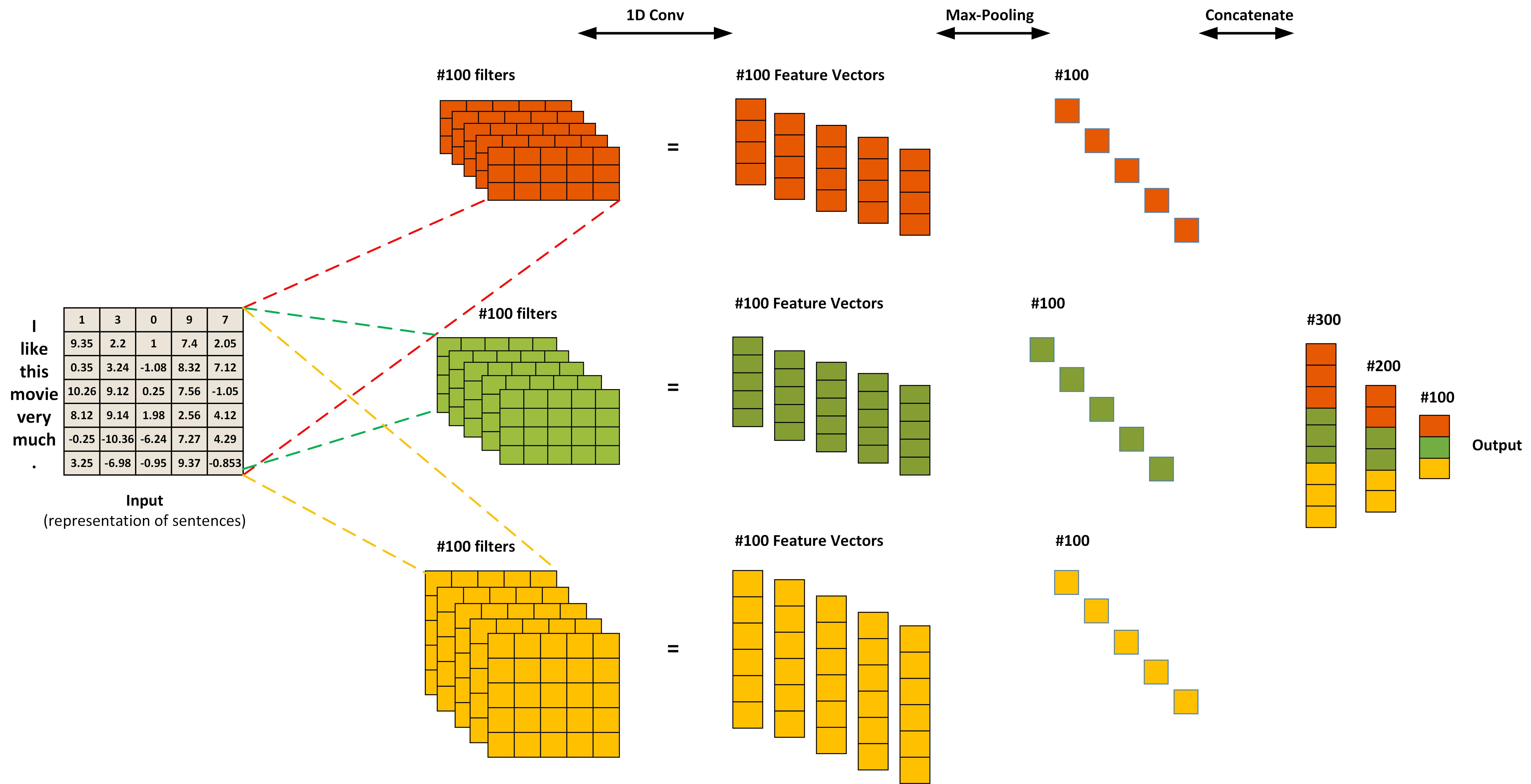}
		\caption{CNN architecture to process the documents that describe items.}
		\label{figure=CNN structure}
	\end{figure}

	As shown in fig \ref{figure=overal struct}, the item section consists of two parts:
	\newline
	\textbf{Document Information Modeling Part}: We use the popular pre-trained word embedding model, GloVe \footnote{\url{https://nlp.stanford.edu/projects/glove/}}, that converts each word to a dense vector with a fixed length. Suppose we have $\rho$ words in a document that describes the item, so we obtain a matrix $D\subset \mathbb{R}^{\rho \times l}$ where $l$ is the size of embedding for each word. As shown in fig\ref{figure=CNN structure}, we use one dimensional CNN with multiple filters to project input to vectors and capture various types of contextual features:
	\begin{eqnarray}
		c_{i}^{j}&=&f(W.x(i:i+h-1)+b) \\
		c^{j}&=&[c_{1}^{j},c_{2}^{j},c_{3}^{j},...,c_{i+h-1}^{j}]
	\end{eqnarray}
	Where $c_{i}^{j} \in \mathbb{R}$, $i$ is the number of convolution operations, $m$ is the number of convolutional kernels, $f$ is the activation function, $W$ is the weight matrix, and $b\in \mathbb{R}$ is the bias.
	\newline 
	As to the variable length of each document, we use the convolution architecture in \citep{collobert2011natural,rakhlin2016convolutional}. After passing the document through the convolution layer, we obtain a feature vector with variable length for each kernel weight. By using max-pooling, each vector is converted to a scalar that extracts the features from the previous layer.
	\begin{eqnarray}
		q=[max(c^{1}),max(c^{2}),...,max(c^{j}),...,max(c^{n})]	
	\end{eqnarray}
	Where $q\subset \mathbb{R}^{n}$ is a fixed-length vector. Finally, we pass the vector through a fully connected layer.
	
	\textbf{User Information Modeling Part}: Like the user modeling section, we represent each item as a set of users interacting with the target item. As we said, each user can express their satisfaction, denoted as rating score, which means that all interactions are not of equal importance. We use this information and, similar to \cite{fan2019graph} combine the rating embedding with the user embedding. By passing each input vector through an attention network, the attention weights are obtained. Next, each vector is multiplied by the corresponding weight.
	\newline
	Finally, we concatenate these two vectors and pass them through a fully connected layer.
	
	\subsection{Rating Prediction Section}
	Finally, the user and item latent vectors are obtained according to the method proposed in the previous sections. Then, they are concatenated and passed through fully connected layers to predict ratings using the following equations:
	\newline
	\begin{eqnarray}
		L_{1}&=&\sigma_{1}(W_{1} \times Concat(U,I)+b_{1}) \nonumber \\
		L_{2}&=&\sigma_{2}(W_{2} \times L_{1}+b_{2}) \\
		\; &\vdots&        \nonumber  \\
		L_{k}&=&\sigma_{k}(W_{k} \times L_{k-1}+b_{k}) \nonumber
	\end{eqnarray}
	In which $k$, $\sigma _{i}$, $W$, and $b$ respectively denote the number of layers, activation function, weights matrix and bias vector. In addition, $U$ and $I$ represent user and item latent vectors.
	\section{Experiment} \label{section:experiment}
	In this section we first introduce the evaluation datasets, some implementation details, optimization and evaluation methods. Then we analyze our model from different aspects.
	\subsection{Experimental Settings}
	\subsubsection{Datasets}
	To demonstrate the effectiveness of our model, we used Movielens\cite{harper2015movielens}\footnote{\url{https://movielens.org/}} and Amazon\footnote{\url{http://jmcauley.ucsd.edu/data/amazon/}} datasets. The datasets contain user ratings of items on a scale of 1 to 5 as Table \ref{Tab1}.
	\begin{table}[h!]
		\caption{Statistics of the evaluation datasets.}
		\label{Tab1}
		\centering
		\begin{tabular}{lllll}
			\toprule
			Dataset     & \# user     & \# item  & \# Interaction &  Density  \\
			\midrule
			ML-1m & 6,040  & 3,544     & 993,482 & 4.641 \% \\
			ML-10m & 69,878 & 10,073 & 9,945,875 & 1.413 \% \\
			Amazon& 29,757& 15,149 & 135,188 & 0.030 \% \\
			\bottomrule
		\end{tabular}
	\end{table}
	\newline
	Since our evaluation datasets do not contain the item's description, we used IMDB\footnote{\url{https://www.imdb.com/}} dataset, which includes the storyline and summary of the movies that was provided by MovieLens and Amazon researchers.
	\subsubsection{Implementation Detail}
	We implemented our model using TensorFlow library\cite{tensorflow2015-whitepaper} in Python and used GeForce GTX 1660Ti GPU for the computations. The datasets were divided into $80\%$ for the training set, $10\%$ for the validation set, and the remaining $10\%$ for the testing set. We have at least one example for each user and item in the training set. In addition, We set the batch size to 256 and removed the movies that do not have descriptions and users with less than three ratings. 
	\newline
	We used the following settings in CNN architecture: 1) Set the word embedding length to 300 and used GloVe pre-trained word embedding model. 2) Set the maximum length of each item's description to 300 words. 3) Removed the stop words. 4) Selected 8000  top words that have the highest TF-IDF scores as vocabulary set. 5) used filters with different sizes (3,4,5) and 100 filters per size to capture semantic information of documents.
	\newline
	We know that each user may have a different rated item vector length, and each item's description may not have the same number of words. We use the masking trick to solve this problem by adding masks to ensure all cases have the same length.
	
	\subsubsection{Optimization and Evaluation Metric}
	To evaluate our model and compare with the baselines, we use Root Mean Squared Error (RMSE), Hit Ratio (HR), and Normalized Discounted Cumulative Gain (NDCG) as evaluation metrics that are defined as:
	\begin{eqnarray}
		RMSE&=&\sqrt{\frac{\sum_{i,j=1}^{N,M}(r_{ij}-\hat{r}_{ij})^{2}}{T}} \\
		HR@K&=&\frac{Number~of~cache~hits}{Number~of~cache~hits+Number~of~cache~Misses} \\
		NDCG@K&=&Z_{k}\sum_{i=1}^{K}\frac{2^{r_{i}}-1}{log_{2}{(i+1)}}
	\end{eqnarray}
	Where $N$, $M$, $\hat{r}_{ij}$ and $T$ respectively denote the number of the users, items, estimated value of the rating, and the total number of the ratings. Also, $Z_{K}$ is the normalizer to ensure the perfect ranking has a value of 1; $r_{i}$ is the graded relevance of item at position $i$ \cite{he2015trirank}. In HR and NDCG, we adopted the leave one out evaluation metric, which holds out the latest interaction of each user as the testing data and uses the remaining interactions for training. We paired each ground truth item in the test set with 99 randomly sampled negative instances. Hence, the task becomes to rank these negative items with the ground truth item for each user. The performance is judged by HR and NDCG.
	\newline
	For training our model we adopted the Adaptive Moment Estimation (Adam) \cite{kingma2014adam} optimization algorithm, where the learning rate is adopted for each parameter.
	\begin{figure}
		\hspace*{-1cm}
		\centering
		\subfloat[]{
			\label{figure=rating-information}
			\includegraphics[width=0.3\textwidth]{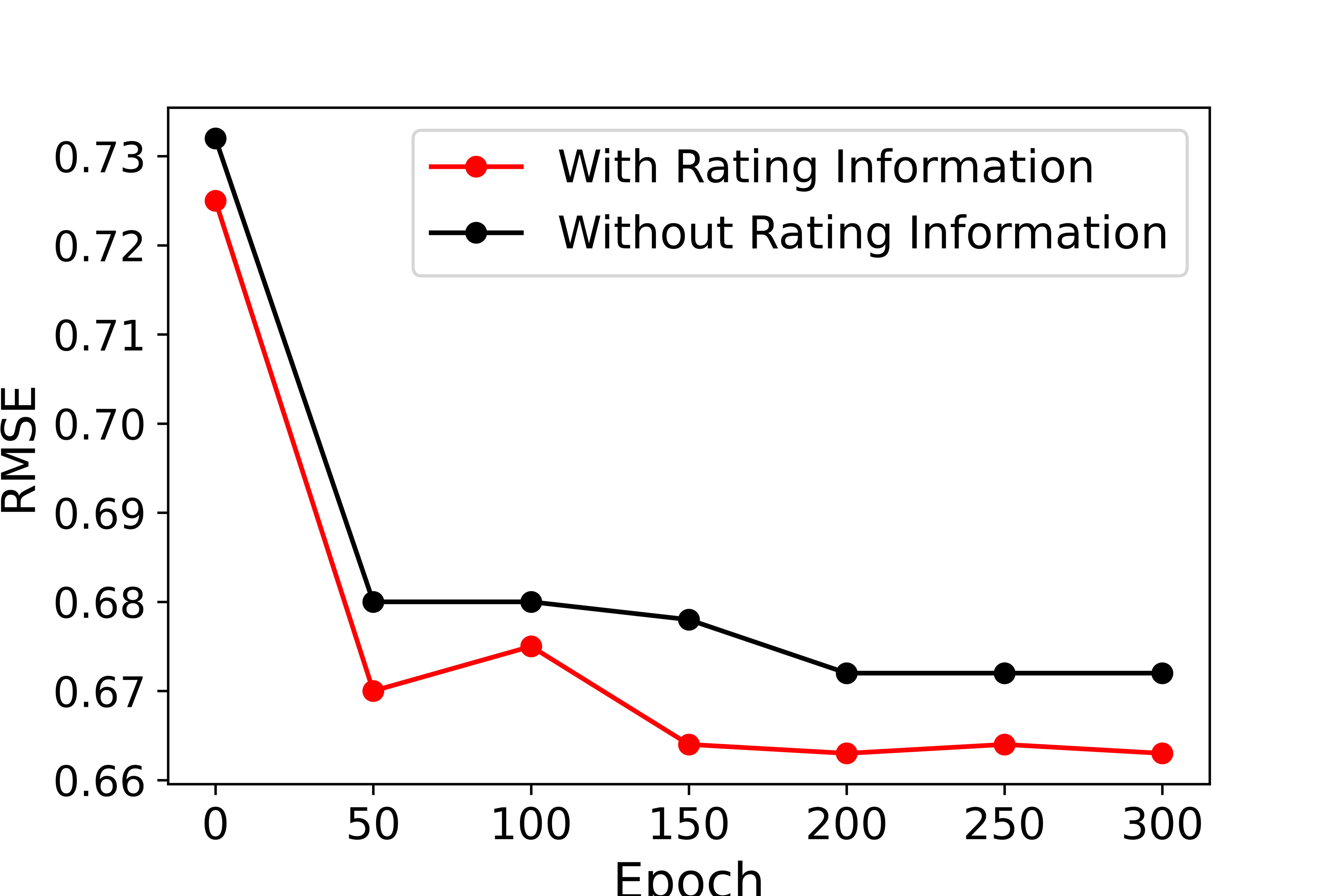}
		}
		\subfloat[]{
			\label{figure=attention-mechanism}
			\includegraphics[width=0.3\textwidth]{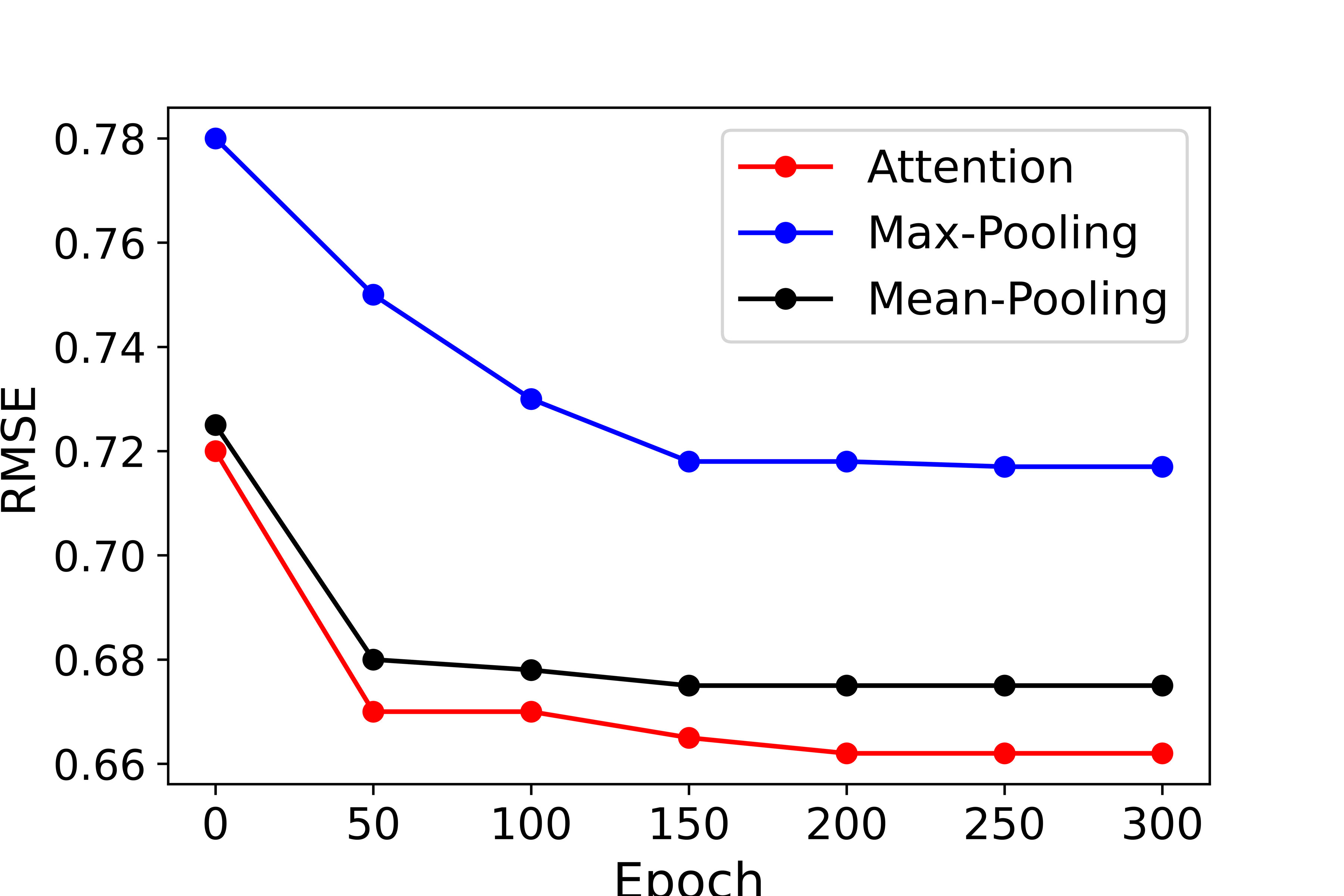}
		}
		\subfloat[]{
			\label{figure=user-information}
			\includegraphics[width=0.3\textwidth]{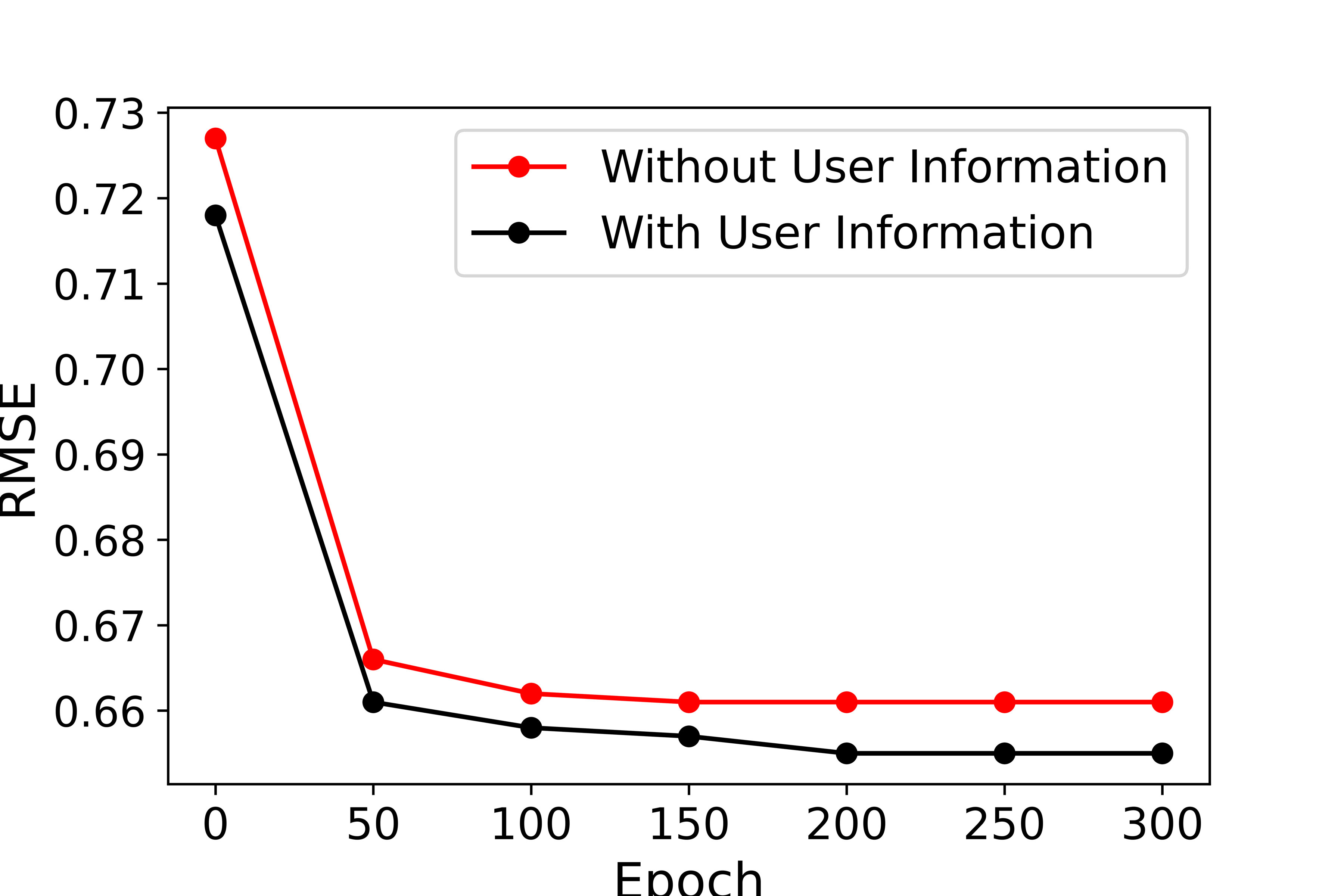}}
		
		\hspace*{-1cm}
		\subfloat[]{
			\label{figure=pre_init_model}
			\includegraphics[width=0.3\textwidth]{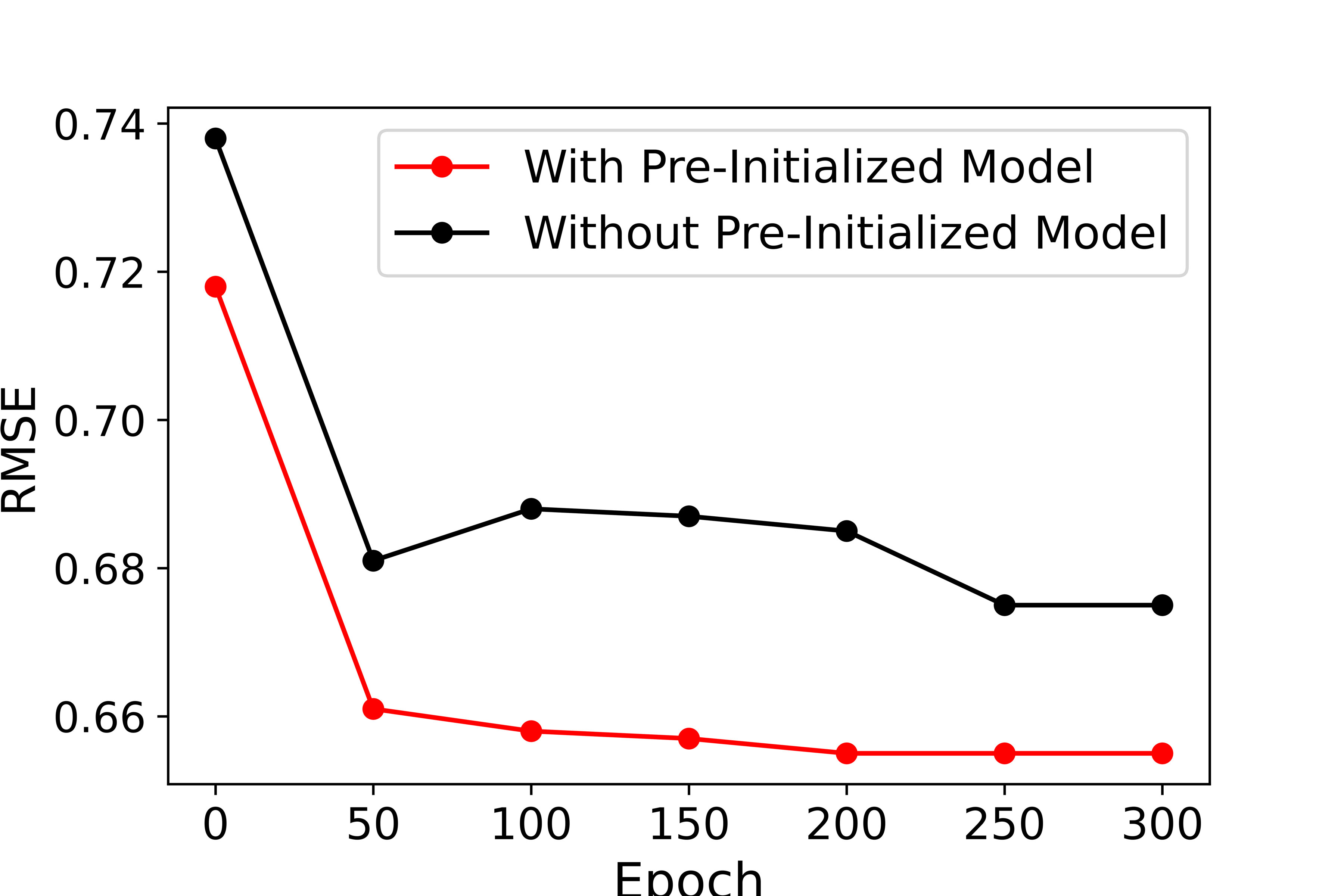}}
		\subfloat[]{
			\label{figure=document_information}
			\includegraphics[width=0.3\textwidth]{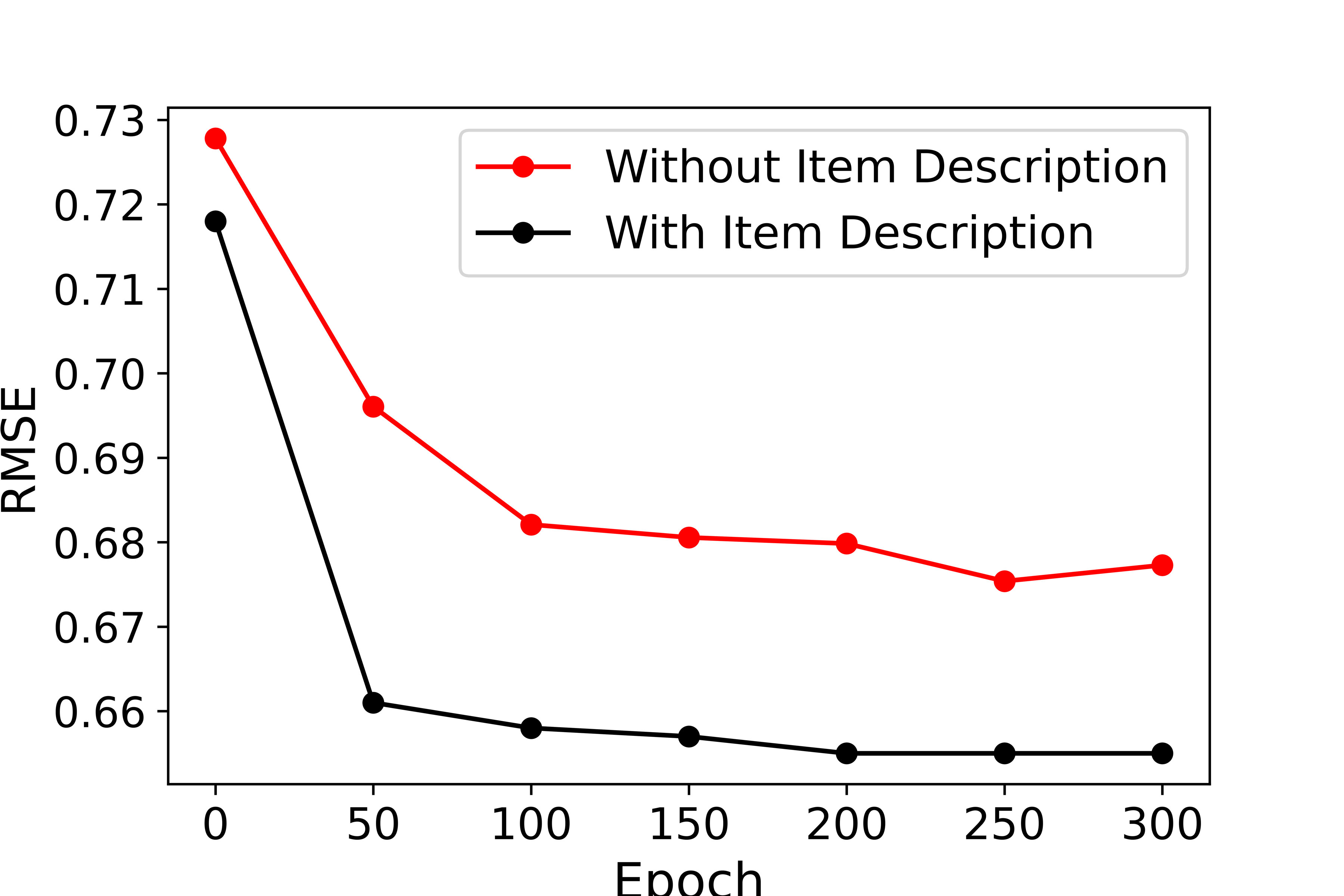}}
		\subfloat[]{
			\label{figure=embedding_size}
			\includegraphics[width=0.3\textwidth]{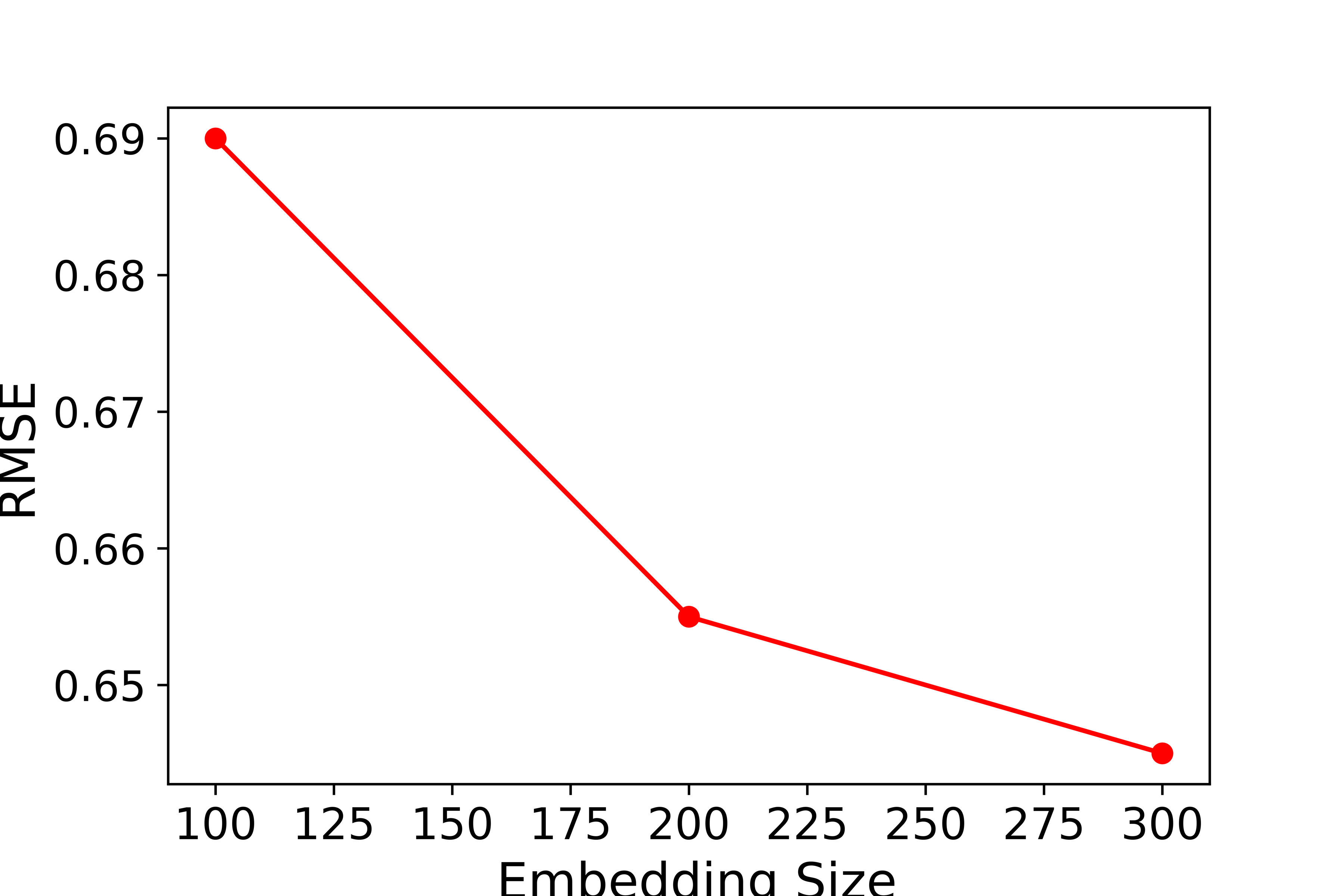}}
		\caption{Effect of rating information, user information, pre-initialized model, document information, and embedding size on ML-1m dataset.}	
	\end{figure}
	\subsubsection{Baselines}
	We compared our model with the following baselines:
	\begin{itemize}
		\item  \textbf{PMF} \cite{mnih2007probabilistic}: Probabilistic Matrix Factorization is a standard rating prediction model which only utilizes user-item rating matrix and models latent factors of users and items by Gaussian distribution.
		\item \textbf{CTR} \cite{wang2011collaborative}: Collaborative Topic Regression is a state-of-the-art model which combines PMF and Latent Dirichlet Allocation (LDA) to predict rating of user $u$ on item $i$.
		\item \textbf{CDL} \cite{wang2015collaborative}: Collaborative Deep Learning is a model that combines auto-encoders and PMF which analyzes documents via SDAE. 
		\item \textbf{Conv MF} \cite{kim2016convolutional}: Convolutional Matrix Factorization is a recent model that uses document information for items as  input and combines PMF and CNN methods to predict rating.
		\item \textbf{NeuMF} \cite{he2017neural}: NeuMF is a state-of-the-art Matrix Factorization model without document information. The original implementation is for ranking task and we adjust its loss to square loss for rating prediction. 
		\item \textbf{Att-ConvCF} \cite{zhang2018integrating}:Attentional ConvCF is a document context aware model that integrates attention mechanism with CF to enhance rating prediction.
		\item \textbf{GNNSR \footnote{Graph Neural Networks for Social Recommendation.}} \cite{fan2019graph}: GNNSR uses graph neural network framework (GraphRec) for social recommendations.
		\item \textbf{RConvCF} \cite{zhang2020extreme}:Residual ConvCF applies residual idea to word embeddings in order to capture semantic information and solve the gradient vanishing problem.
	\end{itemize}
	Among them, PMF, NeuMF and GNNSR are models without document information for rating prediction while other models use document information as auxiliary information for rating prediction.
		\begin{figure}
		\hspace*{-1cm}
		\centering
		\subfloat[ML-1m(RMSE)]{
			\label{figure=RMSE_ML_1m}
			\includegraphics[width=0.33\textwidth]{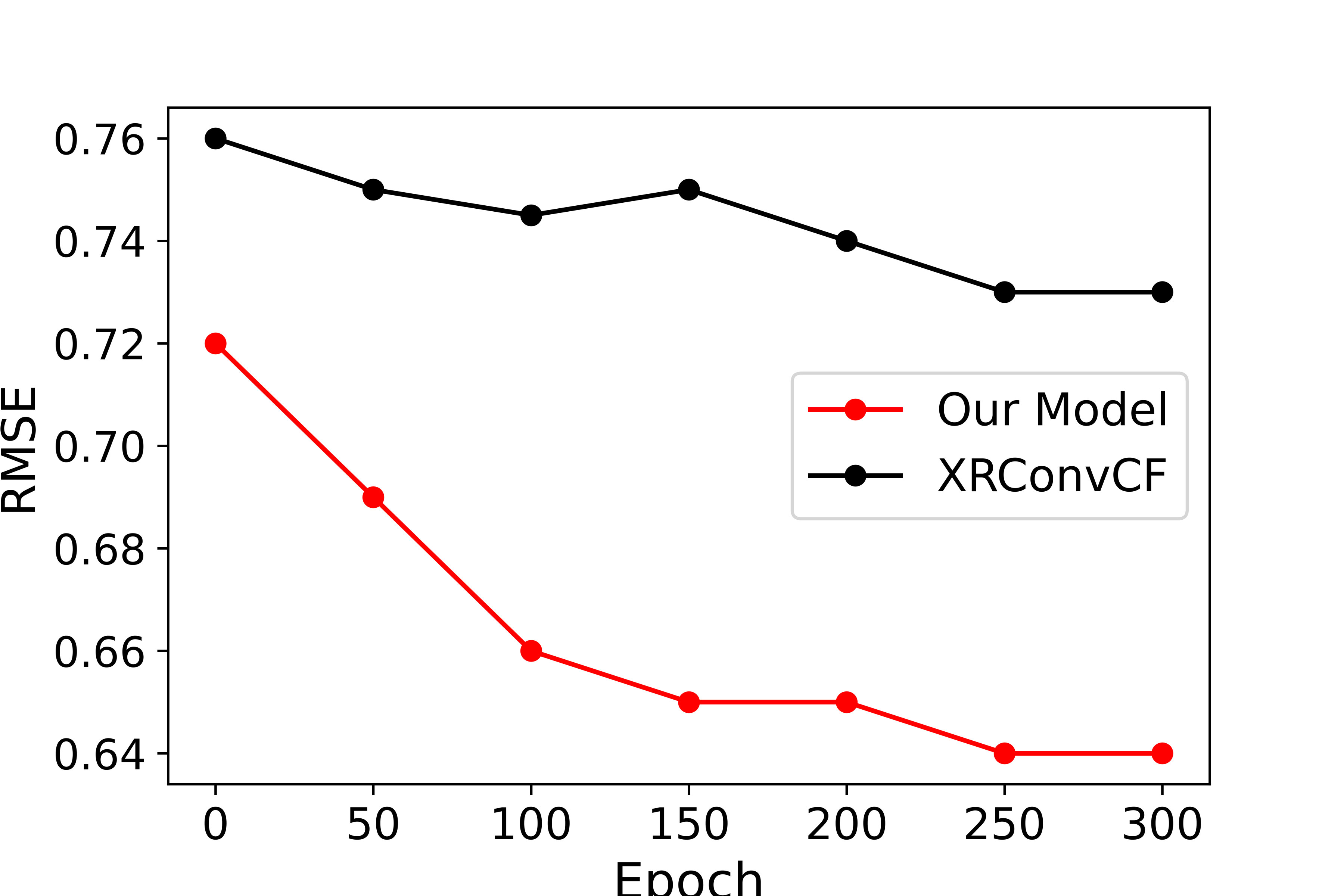}
		}
		\subfloat[ML-10m(RMSE)]{
			\label{figure=RMSE_ML_10m}
			\includegraphics[width=0.33\textwidth]{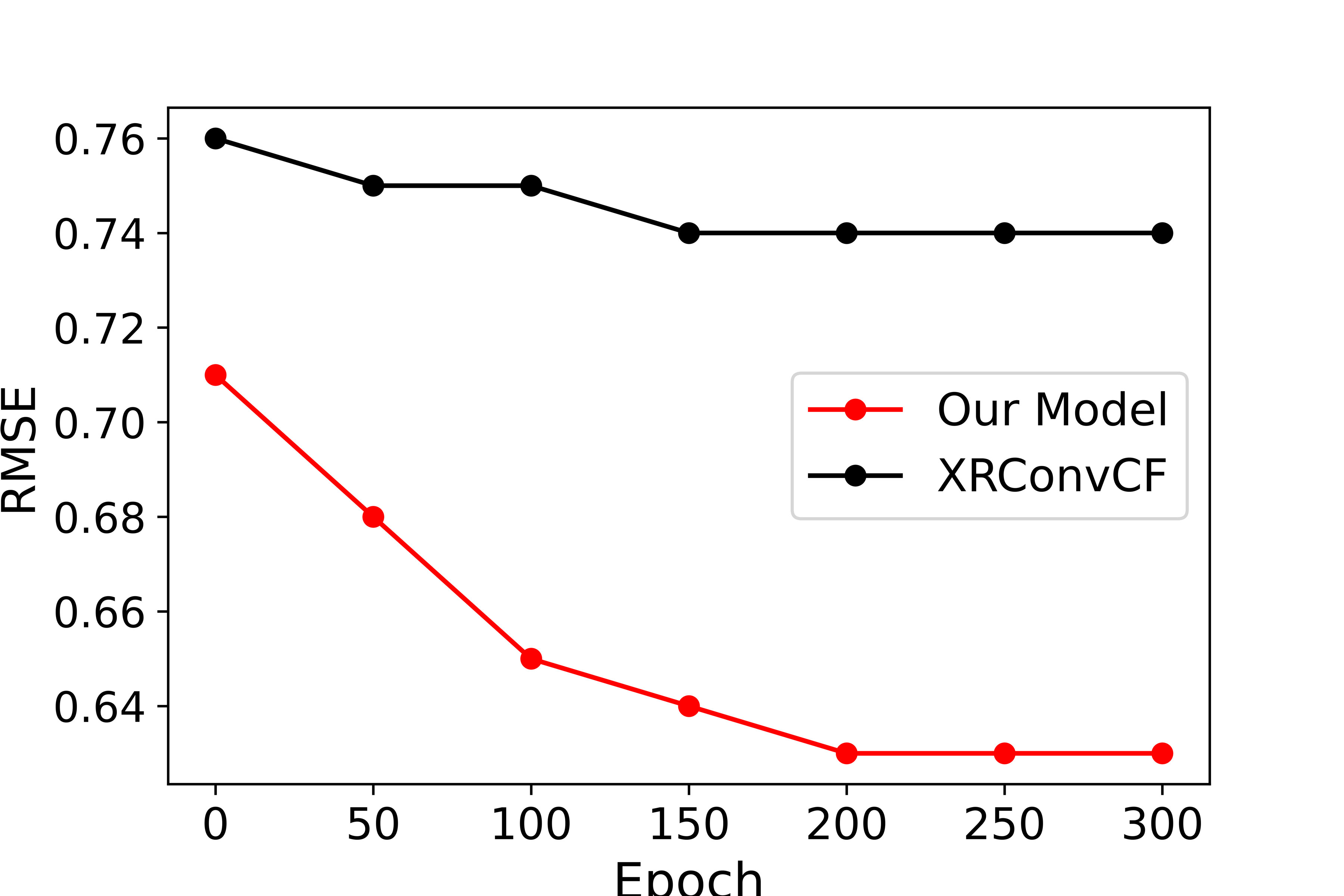}
		}
		\subfloat[Amazon(RMSE)]{
			\label{figure=user-RMSE_AIV}
			\includegraphics[width=0.33\textwidth]{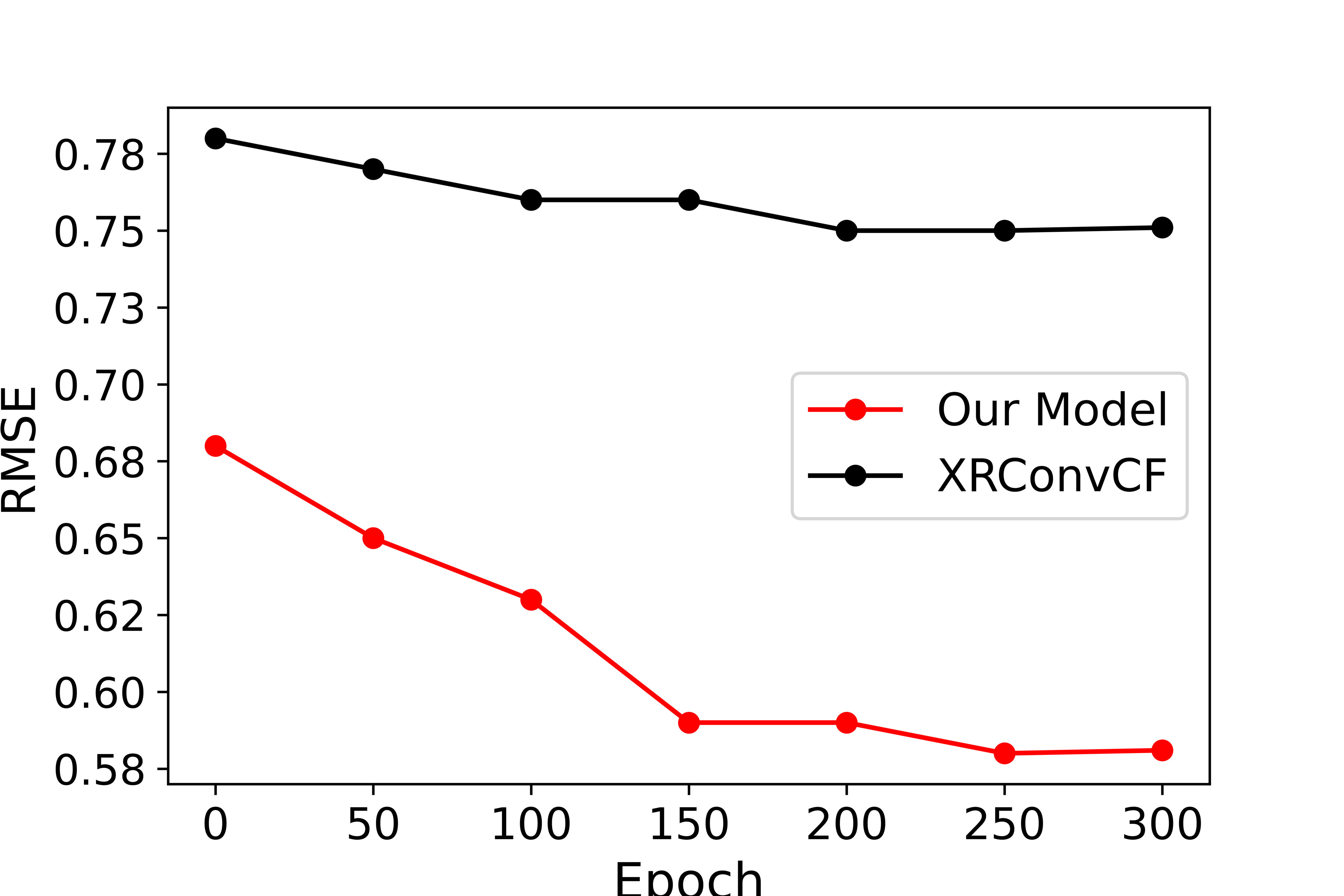}}
		
		\hspace*{-1cm}
		\subfloat[ML-1m(HR@10)]{
			\label{figure=HR_ML_1m}
			\includegraphics[width=0.33\textwidth]{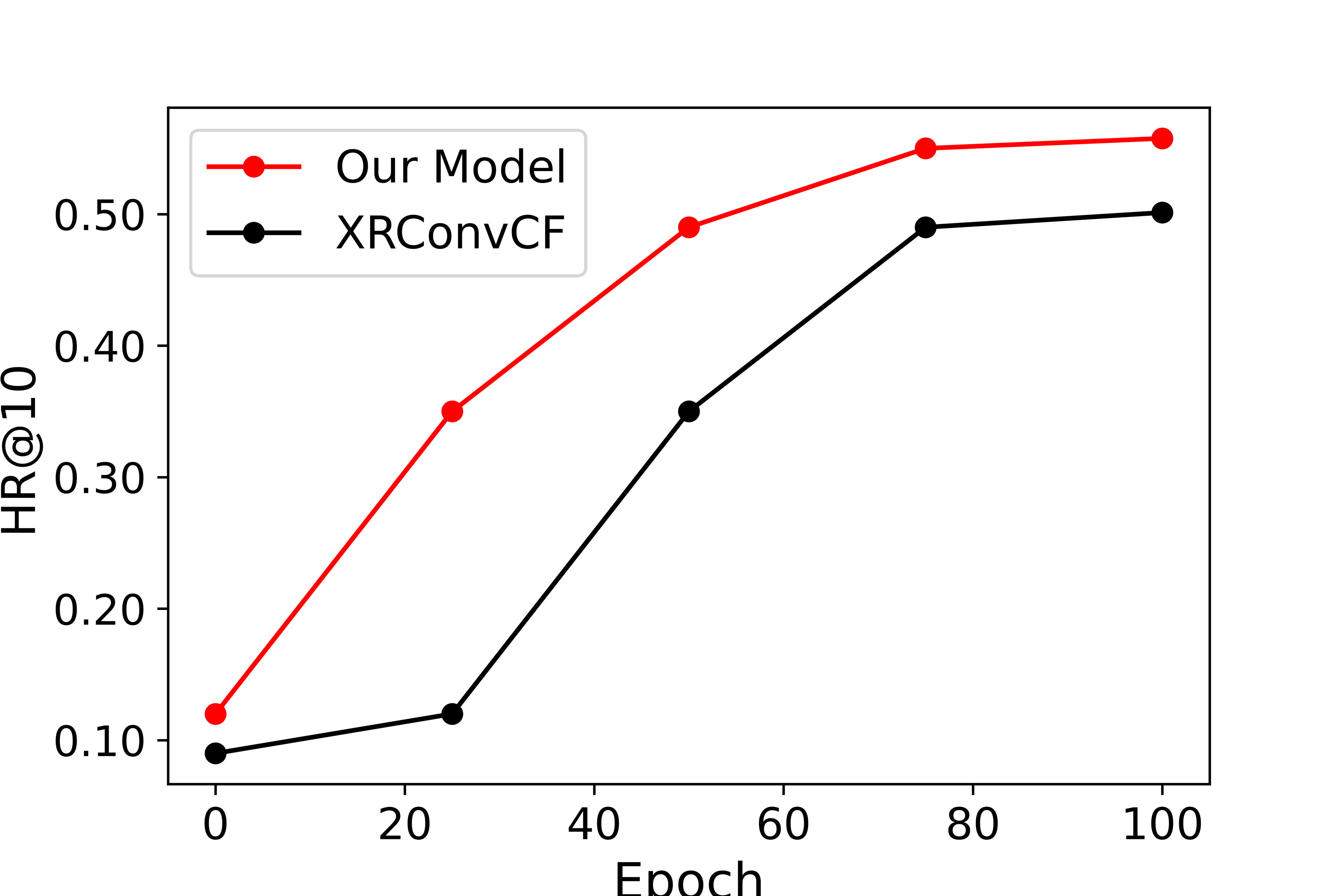}}
		\subfloat[ML-10m(HR@10)]{
			\label{figure=HR_ML_10m}
			\includegraphics[width=0.33\textwidth]{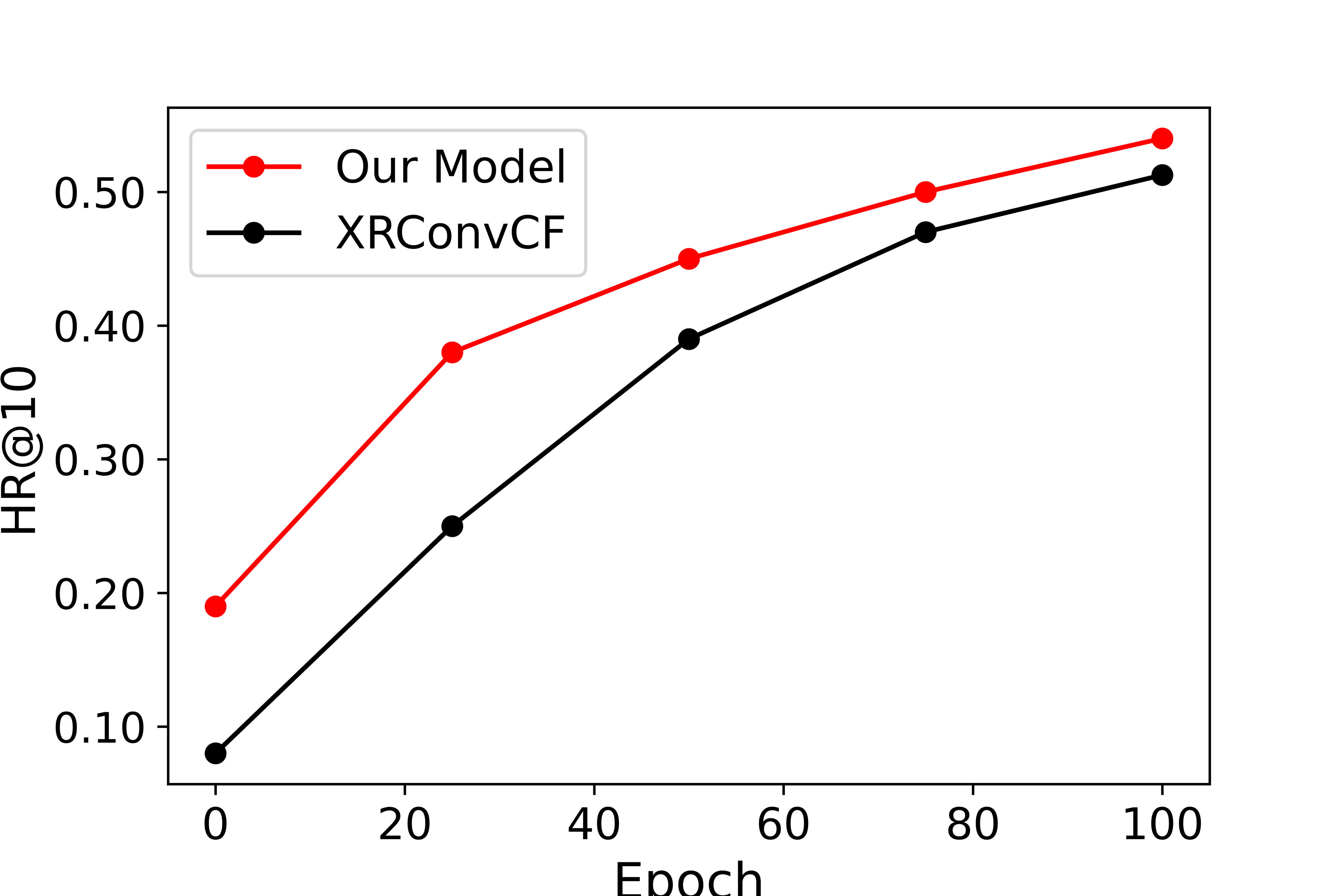}}
		\subfloat[Amazon(HR@10)]{
			\label{figure=HR_AIV}
			\includegraphics[width=0.33\textwidth]{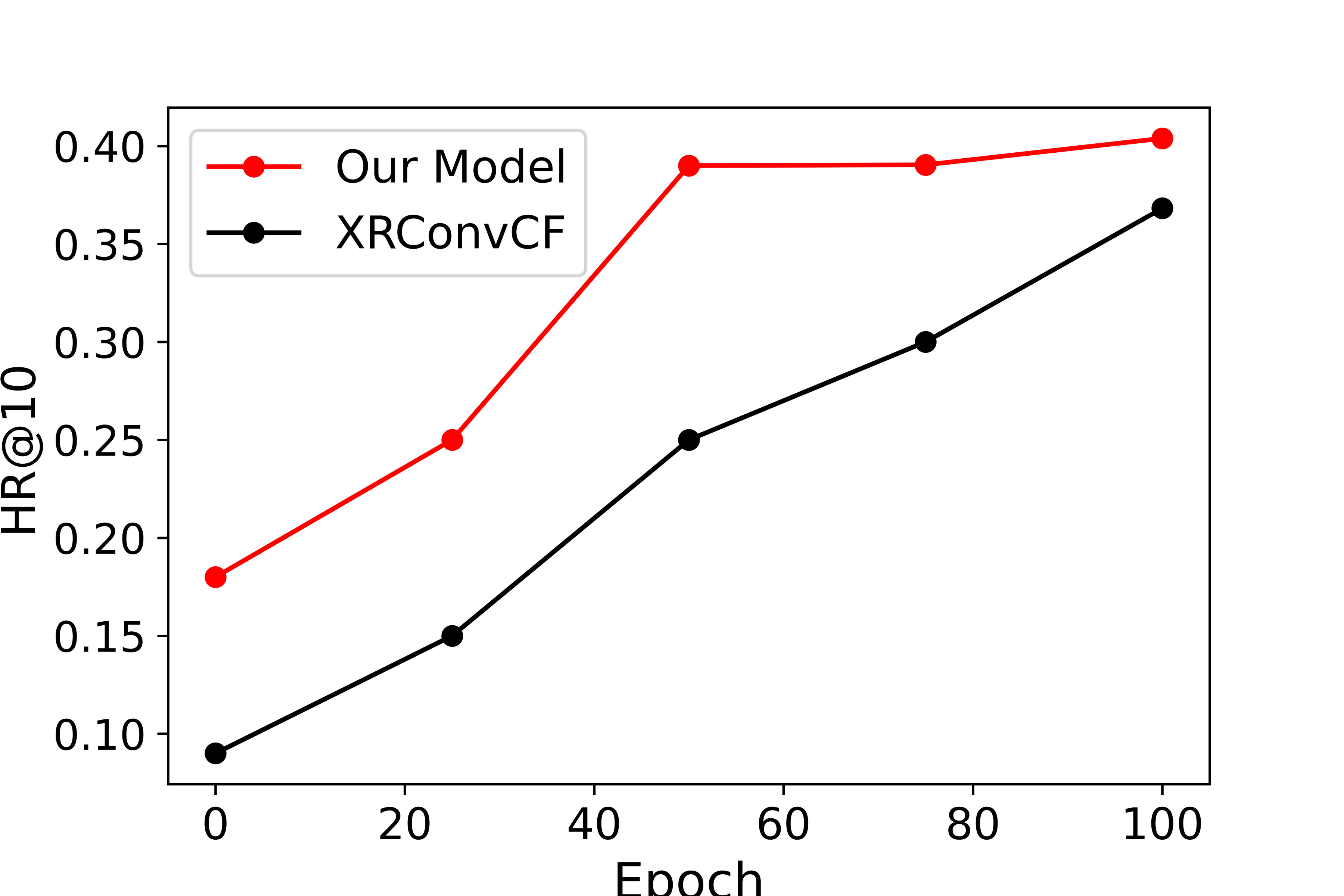}}
		
		\hspace*{-1cm}
		\subfloat[ML-1m(NDCG@10)]{
			\label{figure=NDCG_ML_1m}
			\includegraphics[width=0.33\textwidth]{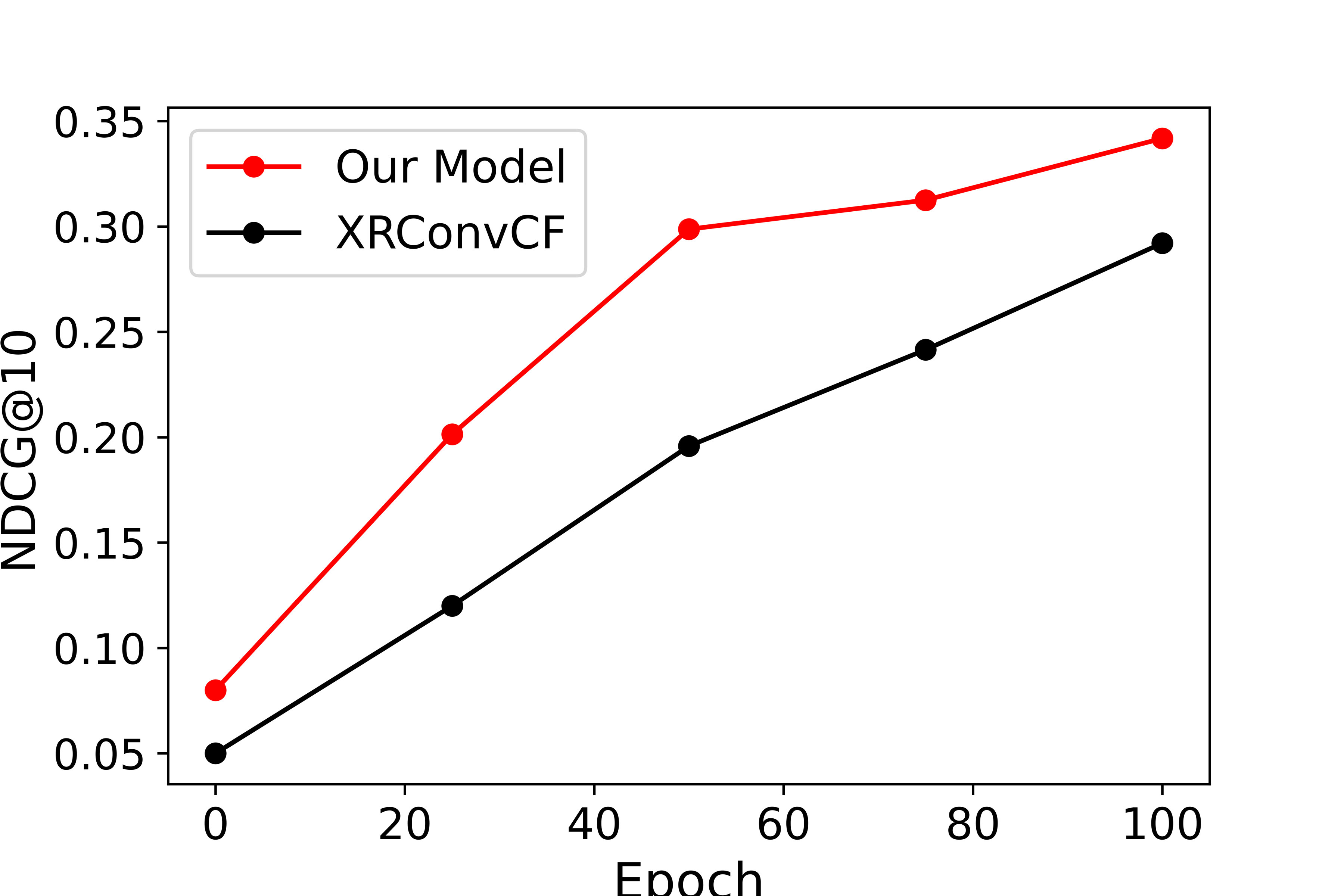}}
		\subfloat[ML-10m(NDCG@10)]{
			\label{figure=NDCG_ML_10m}
			\includegraphics[width=0.33\textwidth]{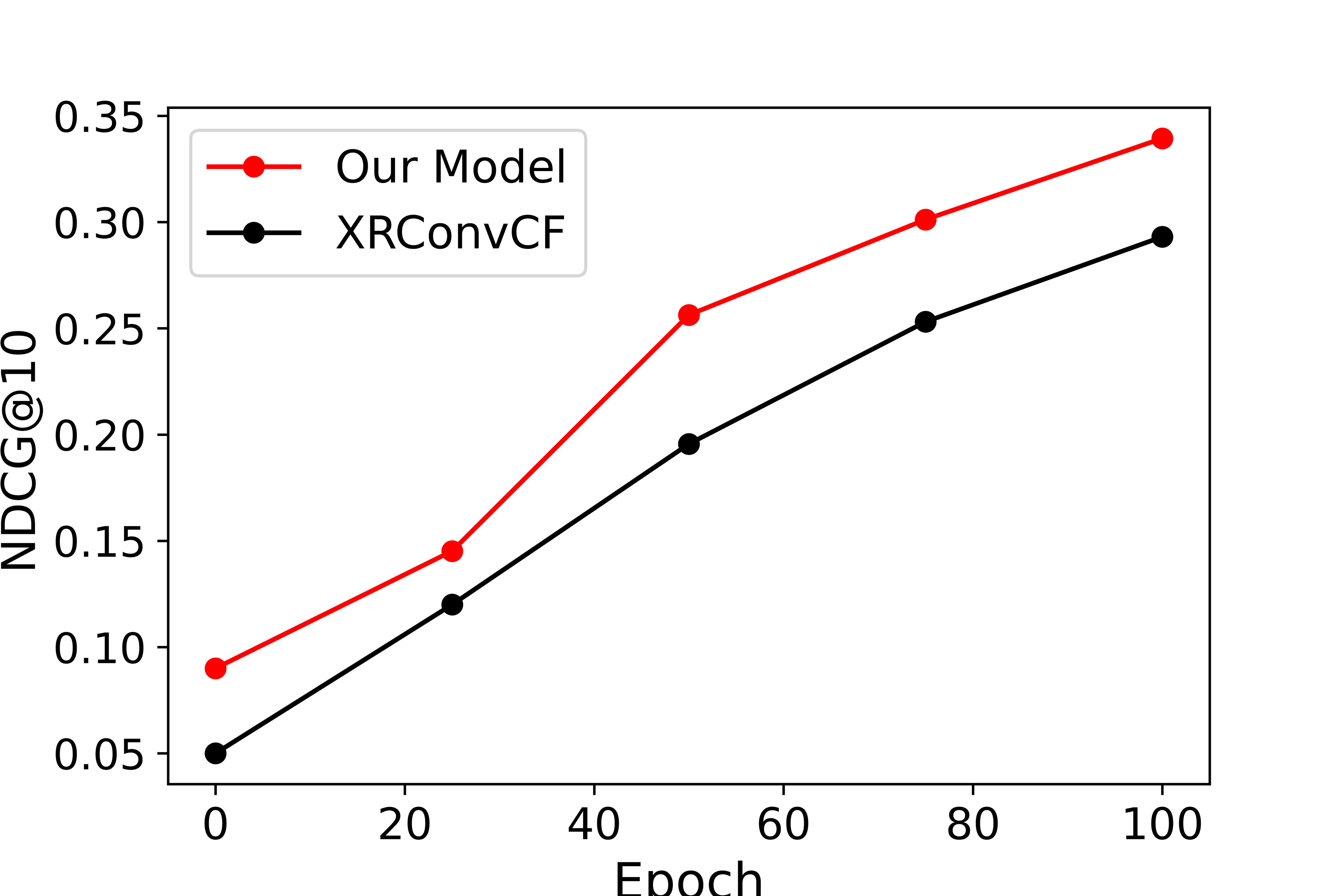}}
		\subfloat[Amazon(NDCG@10)]{
			\label{figure=NDCG_AIV}
			\includegraphics[width=0.33\textwidth]{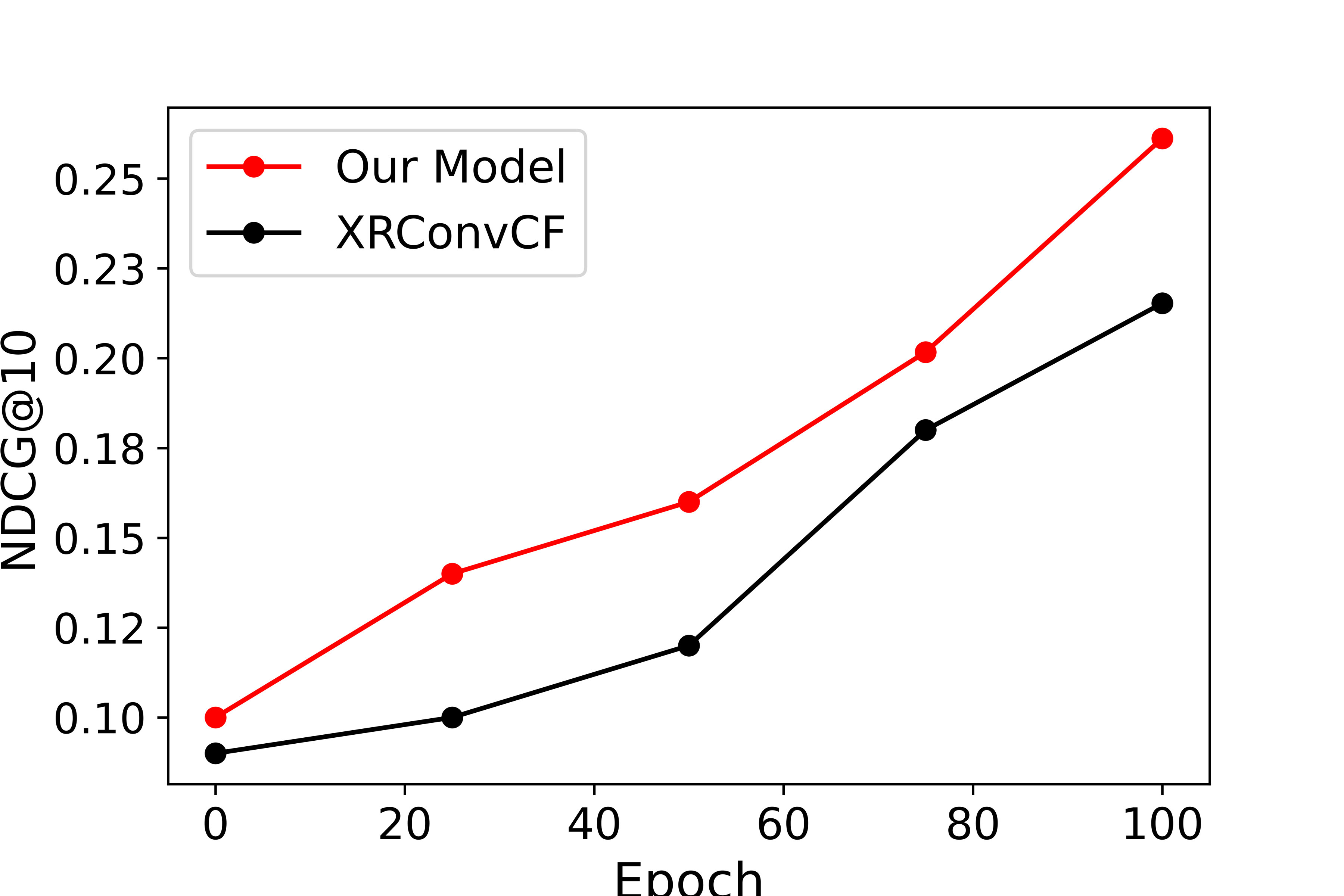}}
		\caption{Testing RMSE loss, HR@10 and NDCG@10 of our proposed model and the best competitor w.r.t. the epochs on ML-1m, ML-10m and Amazon.}
		\label{figure:final_result}	
	\end{figure}
	\subsection{Experimental Results}
	This subsection studies the impact of the rating information, attention mechanism, effect of user information, and word embedding size in item modeling.
	\subsubsection{Effect of the Rating Information}
	A user can express their opinion of the items, which means that all interactions do not have the same importance for the target user. If a user likes the target item very much, they will give it a high score. In this subsection, we compare these two cases and show the results as Fig \ref{figure=rating-information}. It is essential to point out that we didn't consider user information in item modeling for all cases in this subsection. As shown in Fig \ref{figure=rating-information} if we consider the rating information in our model, it achieves better performance.
	
	\subsubsection{Effect of the Attention Mechanism}
	To better understand the proposed model and effectiveness of the involved attention mechanism, we replace it in user and item modeling with Max-Pooling and Mean-Pooling. Fig \ref{figure=attention-mechanism} shows the results on ML-1m. Like the previous subsection, we didn't consider the user information in item modeling.
	As shown in  Fig \ref{figure=attention-mechanism}, The results have met our previous expectation, which means that we have the best performance if we consider attention in our model, while Mean-Pooling achieves better performance than Max-Pooling.

	\begin{table}
	%\sffamily
	
	\caption{Performance Comparison of different methods on ML-1m, ML-10m, and Amazon datasets. Bold Scores are the best, while underlines scores are the second best.}
	%\hspace{-1cm}
	\label{Tab2}
	\centering
	\begin{adjustbox}{width=\columnwidth,center}
	\renewcommand{\arraystretch}{3}
	\begin{tabular}{|l|lll|lll|lll|}
		\hline
		\textbf{Datasets}	&& \textbf{ML-1m} &&& \textbf{ML-10m}&&&\textbf{Amazon}&\\ \hline
		\textbf{Metrics}	& \textbf{HR@10} & \textbf{NDCG@10} & \textbf{RMSE} & \textbf{HR@10}  &   \textbf{NDCG@10}  &  \textbf{RMSE}   & \textbf{HR@10} & \textbf{NDCG@10} & \textbf{RMSE} \\ \hline
		
		\textbf{PMF} &$N/A$&$N/A$&0.897  &$N/A$&$N/A$&0.831&$N/A$&$N/A$&1.412 \\ \hline
		
		\textbf{POP} & 0.136&0.062&$N/A$&0.129&0.040&$N/A$&0.054&0.032&$N/A$ \\ \hline 
		
		\textbf{CTR}  &$N/A$&$N/A$&0.897&$N/A$&$N/A$&0.828&$N/A$&$N/A$&1.550 \\ \hline
		
		\textbf{BPR-MF} &0.430&0.237&$N/A$&0.342&0.170&$N/A$&0.216&0.099&$N/A$ \\ \hline
		
		\textbf{CDL}  &$N/A$&$N/A$&0.888&$N/A$&$N/A$&0.819&$N/A$&$N/A$&1.359\\ \hline
		
		\textbf{NCF}  &0.348&0.164&0.874&0.301&0.135&0.898&0.265&0.070&1.124 \\ \hline
		\textbf{ConvMF} & 0.497&0.296&0.855&0.496&0.295&0.793&0.357&0.236&1.128 \\ \hline
		
		\textbf{Att-ConvCF} &0.501&0.301&0.740&0.493&0.287&0.760&0.398&0.251&0.772 \\ \hline
		
		\textbf{GNNSR}   &\underline{0.510}&\underline{0.306}&0.751&0.501&\underline{0.294}&\underline{0.742}&\underline{0.385}&\underline{0.236}&0.783\\ \hline
		
		\textbf{xRConvCF} &  0.501&0.292&\underline{0.737}&\underline{0.513}&0.293&0.746&0.368&0.215&\underline{0.752} \\ \hline
		
		\textbf{Our-Model} & \textit{\textbf{0.558}}& \textit{\textbf{0.342}}&\textit{\textbf{0.645}}&\textit{\textbf{0.540}}&\textit{\textbf{0.339}}&\textit{\textbf{0.632}}&\textit{\textbf{0.404}}&\textit{\textbf{0.261}}&\textit{\textbf{0.581}} \\ \hline
		\textbf{Improvement}&9.29\% & 11.77\% &12.41\%&5.34\%&15.4\%&14.8\%&4.9\%&10.7\%&22.8\% \\ \hline
\end{tabular}
	\end{adjustbox}
\end{table}

	\subsubsection{Effect of the User Information in the Item Modeling}
	We now focus on analyzing the effectiveness of the user information in item modeling. At first, We consider the item's description as input of the item section and ignore the user information. Then in the following case, we simultaneously consider the user information and item's description as item input which the results are given in Fig \ref{figure=user-information}. The results show that combining users' behavior and textual information can improve the performance of the model.
	
	\subsubsection{Effects of the Document Information in the Item Modeling}
	In Fig \ref{figure=document_information}, We can see that without document information modeling, the performance of rating prediction has deteriorated significantly, and it justifies our assumption that document information modeling on
	item section has information that can help the model to learn each item's latent vector and improve the recommendation performance.
	
	\subsubsection{Effects of the Embedding Size in the Document Information}
	Figure \ref{figure=embedding_size} shows the performance of our models with respect to various word embedding sizes. When we increase embedding size from 100 to 200, RMSE does not boost; setting the value to 300 leads to the best result.
	
	\subsubsection{Effects of the Word Embedding Pre-trained Model}
	In Fig \ref{figure=pre_init_model}, we investigate the impact of GloVe pre-trained word embedding model on our task. GloVe significantly helps the model to reduce RMSE.

	\subsubsection{Performance Comparison}
	Table \ref{Tab2} summarizes the results of all models on three benchmark datasets, where underlined scores are the best competitor result and bold scores are our model's results. The last row shows the improvement of our model to the best baseline. 
	\newline
	Some of the elements in Table \ref{Tab2} are "N/A", which means that the baseline model has optimized just one objective function: prediction error type or ranking performance. The results for another objective are not available. For example, PMF optimizes just rating prediction errors, so HR and NDCG metric results are not available. The original implementations of ConvMf, Att-ConvCf, GNNSR, and xRConvCF are for the rating prediction recommendation task; we adjust their loss to binary cross-entropy loss for ranking purposes. Also, the original implementation of NCF is for the ranking recommendation task, which we adapt its loss to the rating prediction recommendation task.
	\newline
	Among baseline methods, ConvMF, Att-ConvCF, and xRConvCF use items description as auxiliary information, and the rest of them don't use it.
	\newline
	Table \ref{Tab2} shows our model's results, overall rating prediction error (RMSE), HR, and NDCG over ten baselines on three datasets, which shows that the proposed model achieved significant improvement over all the baselines.
	The first row of fig \ref{figure:final_result} shows RMSE values of the test dataset obtained by the two methods, best baseline and our proposed model, during 300 epochs. The second and third rows show HR and NDCG values during 100 epochs respectively.
	
	\section{Conclusion} \label{section:conclusion}

	This paper aimed to handle the sparsity problem and improve recommendation accuracy by considering contextual information and combining it with historical data. In the user section, we built a profile for each user based on their interacted items, and similarly, we created a profile for each item based on the users who have interacted with it. We analyzed the effect of each section on rating prediction and evaluated our model on three real-world datasets, which demonstrated the effectiveness of our model over the state-of-the-art competitors.
	\newline
	In future work, using one of the state-of-the-art models such as NeuMF, we try to find friends for each user according to the user embedding vector. Then we incorporate the information of these trusted friends into user modeling. Another future work would be exploiting bidirectional models such as BERT\cite{devlin2018bert} to provide a representation for each word by jointly conditioning on both left and right context.

	\bibliographystyle{abbrvnat}

	\bibliography{references}

\begin{thebibliography}{39}
\providecommand{\natexlab}[1]{#1}
\providecommand{\url}[1]{\texttt{#1}}
\expandafter\ifx\csname urlstyle\endcsname\relax
  \providecommand{\doi}[1]{doi: #1}\else
  \providecommand{\doi}{doi: \begingroup \urlstyle{rm}\Url}\fi

\bibitem[Abadi et~al.(2015)Abadi, Agarwal, Barham, Brevdo, Chen, Citro,
  Corrado, Davis, Dean, Devin, Ghemawat, Goodfellow, Harp, Irving, Isard, Jia,
  Jozefowicz, Kaiser, Kudlur, Levenberg, Man\'{e}, Monga, Moore, Murray, Olah,
  Schuster, Shlens, Steiner, Sutskever, Talwar, Tucker, Vanhoucke, Vasudevan,
  Vi\'{e}gas, Vinyals, Warden, Wattenberg, Wicke, Yu, and
  Zheng]{tensorflow2015-whitepaper}
M.~Abadi, A.~Agarwal, P.~Barham, E.~Brevdo, Z.~Chen, C.~Citro, G.~S. Corrado,
  A.~Davis, J.~Dean, M.~Devin, S.~Ghemawat, I.~Goodfellow, A.~Harp, G.~Irving,
  M.~Isard, Y.~Jia, R.~Jozefowicz, L.~Kaiser, M.~Kudlur, J.~Levenberg,
  D.~Man\'{e}, R.~Monga, S.~Moore, D.~Murray, C.~Olah, M.~Schuster, J.~Shlens,
  B.~Steiner, I.~Sutskever, K.~Talwar, P.~Tucker, V.~Vanhoucke, V.~Vasudevan,
  F.~Vi\'{e}gas, O.~Vinyals, P.~Warden, M.~Wattenberg, M.~Wicke, Y.~Yu, and
  X.~Zheng.
\newblock {TensorFlow}: Large-scale machine learning on heterogeneous systems,
  2015.
\newblock URL \url{https://www.tensorflow.org/}.
\newblock Software available from tensorflow.org.

\bibitem[Adomavicius and Tuzhilin(2011)]{adomavicius2011context}
G.~Adomavicius and A.~Tuzhilin.
\newblock Context-aware recommender systems.
\newblock In \emph{Recommender systems handbook}, pages 217--253. Springer,
  2011.

\bibitem[Chaudhari et~al.(2019)Chaudhari, Mithal, Polatkan, and
  Ramanath]{chaudhari2019attentive}
S.~Chaudhari, V.~Mithal, G.~Polatkan, and R.~Ramanath.
\newblock An attentive survey of attention models.
\newblock \emph{arXiv preprint arXiv:1904.02874}, 2019.

\bibitem[Cheng et~al.(2016)Cheng, Koc, Harmsen, Shaked, Chandra, Aradhye,
  Anderson, Corrado, Chai, Ispir, et~al.]{cheng2016wide}
H.-T. Cheng, L.~Koc, J.~Harmsen, T.~Shaked, T.~Chandra, H.~Aradhye,
  G.~Anderson, G.~Corrado, W.~Chai, M.~Ispir, et~al.
\newblock Wide \& deep learning for recommender systems.
\newblock In \emph{Proceedings of the 1st workshop on deep learning for
  recommender systems}, pages 7--10, 2016.

\bibitem[Collobert et~al.(2011)Collobert, Weston, Bottou, Karlen, Kavukcuoglu,
  and Kuksa]{collobert2011natural}
R.~Collobert, J.~Weston, L.~Bottou, M.~Karlen, K.~Kavukcuoglu, and P.~Kuksa.
\newblock Natural language processing (almost) from scratch.
\newblock \emph{Journal of machine learning research}, 12\penalty0
  (ARTICLE):\penalty0 2493--2537, 2011.

\bibitem[Deshpande and Karypis(2004)]{deshpande2004item}
M.~Deshpande and G.~Karypis.
\newblock Item-based top-n recommendation algorithms.
\newblock \emph{ACM Transactions on Information Systems (TOIS)}, 22\penalty0
  (1):\penalty0 143--177, 2004.

\bibitem[Devlin et~al.(2018)Devlin, Chang, Lee, and Toutanova]{devlin2018bert}
J.~Devlin, M.-W. Chang, K.~Lee, and K.~Toutanova.
\newblock Bert: Pre-training of deep bidirectional transformers for language
  understanding.
\newblock \emph{arXiv preprint arXiv:1810.04805}, 2018.

\bibitem[Fan et~al.(2019)Fan, Ma, Li, He, Zhao, Tang, and Yin]{fan2019graph}
W.~Fan, Y.~Ma, Q.~Li, Y.~He, E.~Zhao, J.~Tang, and D.~Yin.
\newblock Graph neural networks for social recommendation.
\newblock In \emph{The World Wide Web Conference}, pages 417--426, 2019.

\bibitem[Harper and Konstan(2015)]{harper2015movielens}
F.~M. Harper and J.~A. Konstan.
\newblock The movielens datasets: History and context.
\newblock \emph{Acm transactions on interactive intelligent systems (tiis)},
  5\penalty0 (4):\penalty0 1--19, 2015.

\bibitem[He et~al.(2015)He, Chen, Kan, and Chen]{he2015trirank}
X.~He, T.~Chen, M.-Y. Kan, and X.~Chen.
\newblock Trirank: Review-aware explainable recommendation by modeling aspects.
\newblock In \emph{Proceedings of the 24th ACM International on Conference on
  Information and Knowledge Management}, pages 1661--1670, 2015.

\bibitem[He et~al.(2016)He, Zhang, Kan, and Chua]{he2016fast}
X.~He, H.~Zhang, M.-Y. Kan, and T.-S. Chua.
\newblock Fast matrix factorization for online recommendation with implicit
  feedback.
\newblock In \emph{Proceedings of the 39th International ACM SIGIR conference
  on Research and Development in Information Retrieval}, pages 549--558, 2016.

\bibitem[He et~al.(2017)He, Liao, Zhang, Nie, Hu, and Chua]{he2017neural}
X.~He, L.~Liao, H.~Zhang, L.~Nie, X.~Hu, and T.-S. Chua.
\newblock Neural collaborative filtering.
\newblock In \emph{Proceedings of the 26th international conference on world
  wide web}, pages 173--182, 2017.

\bibitem[He et~al.(2018)He, He, Song, Liu, Jiang, and Chua]{he2018nais}
X.~He, Z.~He, J.~Song, Z.~Liu, Y.-G. Jiang, and T.-S. Chua.
\newblock Nais: Neural attentive item similarity model for recommendation.
\newblock \emph{IEEE Transactions on Knowledge and Data Engineering},
  30\penalty0 (12):\penalty0 2354--2366, 2018.

\bibitem[Hu et~al.(2008)Hu, Koren, and Volinsky]{hu2008collaborative}
Y.~Hu, Y.~Koren, and C.~Volinsky.
\newblock Collaborative filtering for implicit feedback datasets.
\newblock In \emph{2008 Eighth IEEE International Conference on Data Mining},
  pages 263--272. Ieee, 2008.

\bibitem[Kim et~al.(2016)Kim, Park, Oh, Lee, and Yu]{kim2016convolutional}
D.~Kim, C.~Park, J.~Oh, S.~Lee, and H.~Yu.
\newblock Convolutional matrix factorization for document context-aware
  recommendation.
\newblock In \emph{Proceedings of the 10th ACM conference on recommender
  systems}, pages 233--240, 2016.

\bibitem[Kingma and Ba(2014)]{kingma2014adam}
D.~P. Kingma and J.~Ba.
\newblock Adam: A method for stochastic optimization.
\newblock \emph{arXiv preprint arXiv:1412.6980}, 2014.

\bibitem[Koren et~al.(2009)Koren, Bell, and Volinsky]{koren2009matrix}
Y.~Koren, R.~Bell, and C.~Volinsky.
\newblock Matrix factorization techniques for recommender systems.
\newblock \emph{Computer}, 42\penalty0 (8):\penalty0 30--37, 2009.

\bibitem[Ling et~al.(2014)Ling, Lyu, and King]{10.1145/2645710.2645728}
G.~Ling, M.~R. Lyu, and I.~King.
\newblock Ratings meet reviews, a combined approach to recommend.
\newblock In \emph{Proceedings of the 8th ACM Conference on Recommender
  Systems}, RecSys '14, page 105–112, New York, NY, USA, 2014. Association
  for Computing Machinery.
\newblock ISBN 9781450326681.
\newblock \doi{10.1145/2645710.2645728}.
\newblock URL \url{https://doi.org/10.1145/2645710.2645728}.

\bibitem[Livne et~al.(2019)Livne, Unger, Shapira, and Rokach]{livne2019deep}
A.~Livne, M.~Unger, B.~Shapira, and L.~Rokach.
\newblock Deep context-aware recommender system utilizing sequential latent
  context.
\newblock \emph{arXiv preprint arXiv:1909.03999}, 2019.

\bibitem[McAuley and Leskovec(2013)]{mcauley2013hidden}
J.~McAuley and J.~Leskovec.
\newblock Hidden factors and hidden topics: understanding rating dimensions
  with review text.
\newblock In \emph{Proceedings of the 7th ACM conference on Recommender
  systems}, pages 165--172, 2013.

\bibitem[Mnih and Salakhutdinov(2007)]{mnih2007probabilistic}
A.~Mnih and R.~R. Salakhutdinov.
\newblock Probabilistic matrix factorization.
\newblock \emph{Advances in neural information processing systems},
  20:\penalty0 1257--1264, 2007.

\bibitem[Mnih and Salakhutdinov(2008)]{mnih2008probabilistic}
A.~Mnih and R.~R. Salakhutdinov.
\newblock Probabilistic matrix factorization.
\newblock In \emph{Advances in neural information processing systems}, pages
  1257--1264, 2008.

\bibitem[Mu(2018)]{mu2018survey}
R.~Mu.
\newblock A survey of recommender systems based on deep learning.
\newblock \emph{Ieee Access}, 6:\penalty0 69009--69022, 2018.

\bibitem[Nakhli et~al.(2019)Nakhli, Moradi, and Sadeghi]{nakhli2019movie}
R.~E. Nakhli, H.~Moradi, and M.~A. Sadeghi.
\newblock Movie recommender system based on percentage of view.
\newblock In \emph{2019 5th Conference on Knowledge Based Engineering and
  Innovation (KBEI)}, pages 656--660. IEEE, 2019.

\bibitem[Rakhlin(2016)]{rakhlin2016convolutional}
A.~Rakhlin.
\newblock Convolutional neural networks for sentence classification.
\newblock \emph{GitHub}, 2016.

\bibitem[Sarwar et~al.(2001)Sarwar, Karypis, Konstan, and
  Riedl]{sarwar2001item}
B.~Sarwar, G.~Karypis, J.~Konstan, and J.~Riedl.
\newblock Item-based collaborative filtering recommendation algorithms.
\newblock In \emph{Proceedings of the 10th international conference on World
  Wide Web}, pages 285--295, 2001.

\bibitem[Smirnova and Vasile(2017)]{smirnova2017contextual}
E.~Smirnova and F.~Vasile.
\newblock Contextual sequence modeling for recommendation with recurrent neural
  networks.
\newblock In \emph{Proceedings of the 2nd workshop on deep learning for
  recommender systems}, pages 2--9, 2017.

\bibitem[Sun et~al.(2019)Sun, Liu, Wu, Pei, Lin, Ou, and
  Jiang]{sun2019bert4rec}
F.~Sun, J.~Liu, J.~Wu, C.~Pei, X.~Lin, W.~Ou, and P.~Jiang.
\newblock Bert4rec: Sequential recommendation with bidirectional encoder
  representations from transformer.
\newblock In \emph{Proceedings of the 28th ACM international conference on
  information and knowledge management}, pages 1441--1450, 2019.

\bibitem[Utz et~al.(2012)Utz, Kerkhof, and Van Den~Bos]{utz2012consumers}
S.~Utz, P.~Kerkhof, and J.~Van Den~Bos.
\newblock Consumers rule: How consumer reviews influence perceived
  trustworthiness of online stores.
\newblock \emph{Electronic Commerce Research and Applications}, 11\penalty0
  (1):\penalty0 49--58, 2012.

\bibitem[Von~Helversen et~al.(2018)Von~Helversen, Abramczuk, Kope{\'c}, and
  Nielek]{von2018influence}
B.~Von~Helversen, K.~Abramczuk, W.~Kope{\'c}, and R.~Nielek.
\newblock Influence of consumer reviews on online purchasing decisions in older
  and younger adults.
\newblock \emph{Decision Support Systems}, 113:\penalty0 1--10, 2018.

\bibitem[Wang and Blei(2011)]{wang2011collaborative}
C.~Wang and D.~M. Blei.
\newblock Collaborative topic modeling for recommending scientific articles.
\newblock In \emph{Proceedings of the 17th ACM SIGKDD international conference
  on Knowledge discovery and data mining}, pages 448--456, 2011.

\bibitem[Wang et~al.(2015)Wang, Wang, and Yeung]{wang2015collaborative}
H.~Wang, N.~Wang, and D.-Y. Yeung.
\newblock Collaborative deep learning for recommender systems.
\newblock In \emph{Proceedings of the 21th ACM SIGKDD international conference
  on knowledge discovery and data mining}, pages 1235--1244, 2015.

\bibitem[Wei et~al.(2017)Wei, He, Chen, Zhou, and Tang]{wei2017collaborative}
J.~Wei, J.~He, K.~Chen, Y.~Zhou, and Z.~Tang.
\newblock Collaborative filtering and deep learning based recommendation system
  for cold start items.
\newblock \emph{Expert Systems with Applications}, 69:\penalty0 29--39, 2017.

\bibitem[Xiao et~al.(2017)Xiao, Ye, He, Zhang, Wu, and
  Chua]{xiao2017attentional}
J.~Xiao, H.~Ye, X.~He, H.~Zhang, F.~Wu, and T.-S. Chua.
\newblock Attentional factorization machines: Learning the weight of feature
  interactions via attention networks.
\newblock \emph{arXiv preprint arXiv:1708.04617}, 2017.

\bibitem[Xin et~al.(2019)Xin, Chen, He, Wang, Ding, and Jose]{xin2019cfm}
X.~Xin, B.~Chen, X.~He, D.~Wang, Y.~Ding, and J.~Jose.
\newblock Cfm: Convolutional factorization machines for context-aware
  recommendation.
\newblock In \emph{IJCAI}, volume~19, pages 3926--3932, 2019.

\bibitem[Xue et~al.(2019)Xue, He, Wang, Xu, Liu, and Hong]{xue2019deep}
F.~Xue, X.~He, X.~Wang, J.~Xu, K.~Liu, and R.~Hong.
\newblock Deep item-based collaborative filtering for top-n recommendation.
\newblock \emph{ACM Transactions on Information Systems (TOIS)}, 37\penalty0
  (3):\penalty0 1--25, 2019.

\bibitem[Zhang et~al.(2018)Zhang, Zhang, Sun, Feng, and
  He]{zhang2018integrating}
B.~Zhang, H.~Zhang, X.~Sun, G.~Feng, and C.~He.
\newblock Integrating an attention mechanism and convolution collaborative
  filtering for document context-aware rating prediction.
\newblock \emph{IEEE Access}, 7:\penalty0 3826--3835, 2018.

\bibitem[Zhang et~al.(2020)Zhang, Zhu, Yu, Pu, and Feng]{zhang2020extreme}
B.~Zhang, M.~Zhu, M.~Yu, D.~Pu, and G.~Feng.
\newblock Extreme residual connected convolution-based collaborative filtering
  for document context-aware rating prediction.
\newblock \emph{IEEE Access}, 8:\penalty0 53604--53613, 2020.

\bibitem[Zhang et~al.(2019)Zhang, Yao, Sun, and Tay]{zhang2019deep}
S.~Zhang, L.~Yao, A.~Sun, and Y.~Tay.
\newblock Deep learning based recommender system: A survey and new
  perspectives.
\newblock \emph{ACM Computing Surveys (CSUR)}, 52\penalty0 (1):\penalty0 1--38,
  2019.

\end{thebibliography}

\end{document}